\documentclass[
aps,
prd,
10pt,
showpacs,
amsmath,
amssymb,
twocolumn,
nofootinbib,
nobibnotes,
preprintnumbers
]{revtex4-1}
\usepackage{float}
\usepackage{mathrsfs,amsfonts}
\usepackage{mathtools}
\usepackage{tensor}
\usepackage{nicefrac}
\usepackage{url}
\usepackage[hyperindex,breaklinks]{hyperref}
\begin{document}
\title{One-loop divergences for $f(R)$ gravity}
\author{Michael S. Ruf}
\author{Christian F. Steinwachs}
\email{christian.steinwachs@physik.uni-freiburg.de}
\affiliation{Physikalisches Institut, Albert-Ludwigs-Universit\"at Freiburg,\\
Hermann-Herder-Stra\ss e~3, 79104 Freiburg, Germany}
%

\begin{abstract}
We calculate the divergent part of the one-loop effective action for $f(R)$ gravity on an arbitrary background manifold. Our result generalizes previous results for quantum corrections in $f(R)$ gravity, which have been limited to spaces of constant curvature. We discuss a new technical aspect connected to operators with degenerate principal symbol.
Our result has important applications in cosmology and allows to study the quantum equivalence between $f(R)$ theories and scalar-tensor theories.  
\end{abstract}

%
%
\maketitle								  
%
%
%


\section{Introduction}
Together with scalar-tensor theories, $f(R)$ theories provide the most important cosmological models for the early and late time acceleration of the Universe \cite{Sotiriou:2008rp,DeFelice:2010aj,Nojiri:2010wj,Clifton:2011jh,Nojiri:2017ncd}. 

As any modification of general relativity with higher curvature invariants, also $f(R)$ gravity involves higher-derivatives \cite{Stelle:1976gc,Stelle:1977ry,Barth:1983hb}.
While fourth-order gravity, which takes into account all quadratic curvature invariants, is perturbatively renormalizable, it suffers from problems with unitarity due to the appearance of a massive spin-two ghost in the spectrum \cite{Stelle:1976gc}.
In contrast, $f(R)$ theories only propagate the additional scalar degree of freedom---the ``scalaron'' \cite{Starobinsky:1980te}.  
Therefore, $f(R)$ theories avoid the classical Ostrogradski instability and the associated problem with unitarity violation at the quantum level \cite{Woodard:2006nt}---at least within a truncated effective field theory framework.

The calculation of the quantum effective action for theories of gravity has important applications in cosmological models.
In particular, knowledge of the divergent part of the effective action allows to study the renormalization group (RG) improvement of these models.
In the context of general scalar-tensor theories, the one-loop divergences have been calculated in \cite{Barvinsky:1993zg, Shapiro:1995yc, Steinwachs:2011zs}.
These results are important for the RG properties of inflationary models such as nonminimal Higgs inflation \cite{Bezrukov:2007ep,Barvinsky:2008ia,DeSimone:2008ei,Bezrukov:2008ej,Bezrukov:2009db,Barvinsky:2009fy,Barvinsky:2009ii,Bezrukov:2010jz}, whose inflationary predictions for a large nonminimal coupling are almost indistinguishable from Starobinsky's quadratic $f(R)$ model of inflation \cite{Starobinsky:1980te,Bezrukov:2011gp,Kehagias:2013mya}.
This is a particular manifestation of the fact that $f(R)$ theories admit an equivalent scalar-tensor theory formulation at the classical level.
This equivalence can be probed at the quantum level, similarly to the analysis for different parametrizations of scalar-tensor theories \cite{Kamenshchik:2014waa}.

Perturbative calculations of ultraviolet (UV) divergences in theories of gravity have a quite long history, starting with the one-loop result for Einstein gravity, minimally coupled to a free scalar field \cite{tHooft:1974toh}.
The result was extended to Einstein gravity with a cosmological constant in \cite{Christensen:1979iy} and to Einstein gravity at the two-loop level in \cite{Goroff:1985sz, vandeVen:1991gw}.
The one-loop divergences for renormalizable higher-derivative gravity were calculated in \cite{Fradkin:1981iu, Avramidi:1985ki}.
The one-loop effective action for $f(R)$ theories on a de Sitter background has been calculated in \cite{Cognola:2005de}. 
In this paper we generalize the calculation for the one-loop divergences in $f(R)$ gravity to an arbitrary background.

A nonperturbative approach to quantum gravity is the asymptotic safety program, initiated in \cite{Weinberg:1976xy,Weinberg:1980gg}.
The main tool to test the asymptotic safety conjecture is the effective averaged action, which satisfies an exact functional renormalization group equation (ERGE) \cite{Wetterich:1992yh, Reuter:1993kw, Morris:1993qb}.
Practical calculations, however, are limited to truncations restricting the set of operators in the effective averaged action.
Various $f(R)$ truncations of increasing complexity, up to closed flow equations for $f(R)$ gravity, have been obtained \cite{Reuter:1996cp,Reuter:2001ag,Lauscher:2002sq,Codello:2007bd,Machado:2007ea,Benedetti:2009rx,Ohta:2015efa,Ohta:2015fcu, Falls:2016msz,Falls:2016wsa}.
So far, calculations for a general function $f(R)$, have been limited to spaces of constant curvature. 
Since these calculations share many technical aspects of perturbative one-loop calculations, the method for general backgrounds, introduced in this paper, might also find applications in the context of the ERGE.

The paper is structured as follows.
In Sec. \ref{Sec:fRGravity}, we comment on the special structure of the principal symbol, outline our strategy of calculation, perform the gauge-fixing and derive the fluctuation operator and ghost operator. 
In Sec. \ref{Sec:CalcOneLoopDiv}, we show that the calculation of the one-loop divergences reduces to the evaluation of the three functional traces: a standard tensor trace, a standard vector trace and non-standard scalar trace.
In Sec. \ref{Sec:OneLoopDivOnGenBackFinalResult}, we present our main result for the one-loop divergences and its on shell reduction.
In Sec. \ref{Sec:fRdeSitter}, we check our result by independent calculations on a de Sitter background.
In Sec. \ref{Sec:CheckAndComp}, we compare our result with previous calculations in $f(R)$ gravity.
Finally, in Sec. \ref{Sec:Conclusion}, we summarize our results and give a brief outlook on possible applications.

Technical details are provided in several appendixes.
In Appendix \ref{Sec:NotationFormalism} we introduce our notations, the general formalism, calculational tools and a collection of universal functional traces. 
Appendix \ref{AppCoeffBX} contains the tensorial coefficients of two differential operators. In Appendix \ref{App:RepFinRes}, we provide a set of integration by parts identities and present the final result in different bases. 
Finally, in Appendix \ref{TracesIrredDec} we collect results for traces over symmetric transversal-traceless tensors and transverse vectors.


\section{$f(R)$ gravity on arbitrary backgrounds}\label{Sec:fRGravity}
The Euclidean action functional for $f(R)$ gravity in four dimensions reads
\begin{align}
S[g]=\int\mathop{}\!\mathrm{d}^{4}x\, g^{\nicefrac{1}{2}}\,f(R).\label{fRAct}
\end{align}
The linear metric perturbations $\tensor{h}{_\mu_\nu}$ around a fixed but arbitrary background $\tensor{\bar{g}}{_\mu_\nu}$ are defined by
\begin{align}
\tensor{h}{_\mu_\nu}\coloneqq\delta_{g}\tensor{g}{_\mu_\nu}=\tensor{g}{_\mu_\nu}-\tensor{\bar{g}}{_\mu_\nu}.
\end{align}
In what follows, we omit the bars over background quantities.
We denote derivatives of the function $f$ with respect to its argument by a subindex 
\begin{align}
f_{n}\coloneqq\frac{\partial^{n}\!f(R)}{\partial R^n},\quad n\in\mathbb{N}\,.
\end{align}

\subsection{Equations of motion}
For the first variation of the action \eqref{fRAct} we find
\begin{align}
\delta_{g}S[g]={}&\int\mathop{}\!\mathrm{d}^{4}x\, g^{\nicefrac{1}{2}}\,\left[f_{1}\,\delta_{g}R+ g^{\nicefrac{-1}{2}} f\,(\delta_{g} g^{\nicefrac{1}{2}})\right]\,.
\end{align}
The extremal tensor is defined as
\begin{align}
\tensor{\mathscr{E}}{^\mu^\nu}\coloneqq{}& g^{-\nicefrac{1}{2}}\,\frac{\delta S[g]}{\delta \tensor{g}{_\mu_\nu}}\nonumber\\
={}&-\tensor{(f_1)}{_{;\alpha}^{\alpha}}\,\tensor{g}{^\mu^\nu}+\tensor{(f_1)}{^{;\mu\nu}}-f_1\tensor*{R}{^\mu^\nu}+\frac{1}{2}f\tensor{g}{^\mu^\nu}\,.\label{EOM}
\end{align}
We reserve the ``semicolon-postfix'' notation to indicate that covariant derivatives only act on the object they are attached to. In contrast, the ``prefix'' notation indicates that derivatives act on everything to their right. 
The classical equations of motion for the background (``on shell'' condition) are satisfied, if $\tensor{\mathscr{E}}{^\mu^\nu}=0$.
The trace of the extremal is defined as
\begin{align}
\mathscr{E}\coloneqq \tensor{g}{^\mu^\nu}\mathscr{E}_{\mu\nu}=-3\,\tensor{(f_1)}{_{;\alpha}^{\alpha}}-Rf_1+2f\,.\label{OnShellScalar}
\end{align}
The invariance of the action \eqref{fRAct} under diffeomorphisms implies $\tensor{\mathscr{E}}{^\mu^\nu_{;\mu}}=0$.
It is natural to define the rescaled extremal tensor $\tensor{E}{_\mu_\nu}$ and its trace $E$, which are homogeneous functions of degree zero in $f$ and its derivatives $f_n$,
\begin{align}
\tensor{E}{_\mu_\nu}\coloneqq\frac{\tensor{\mathscr{E}}{_\mu_\nu}}{f_1},\quad E\coloneqq\tensor{g}{^\mu^\nu}\tensor{E}{_\mu_\nu}=\frac{\mathscr{E}}{f_1}\,.
\end{align}

\subsection{Hessian and degeneracy of the principal symbol}\label{SubSec:DegPrincipalPart}
The second variation of the action \eqref{fRAct} can be written in the suggestive form
\begin{align}
&\delta^2_{g} S[g]=\int\mathop{}\!\mathrm{d}^{4}{x}\, g^{\nicefrac{1}{2}}\left\{f_2(\delta_g R)^2\phantom{ (\delta_g^2 g^{\nicefrac{1}{2}})}\right.\nonumber\\
&\left.+f_1\left[\delta_g^2 R+2 g^{\nicefrac{-1}{2}}(\delta_g g^{\nicefrac{1}{2}})(\delta_g R)+R\, g^{\nicefrac{-1}{2}} (\delta_g^2 g^{\nicefrac{1}{2}})\right]\right.\nonumber\\
&\left.+(f-Rf_1)\, g^{\nicefrac{-1}{2}} (\delta_g^2 g^{\nicefrac{1}{2}})\right\}\,.\label{SecVarAct}
\end{align}
From \eqref{SecVarAct}, it is obvious that the Hessian in $f(R)$ gravity leads to a fourth-order operator (for $f_{2}\neq0$), as $f(R)$ is a function of the undifferentiated Ricci scalar $R$ only.
The special structure of the second variation shows that all fourth-order derivatives are included in the $f_2(\delta_g R)^2$ term in the first line.
Up to the overall factor $f_1$, the expression in the second line resembles the second variation of the Einstein-Hilbert action.
The last line reduces to a cosmological constant in the Einstein-Hilbert case\footnote{The overall sign in \eqref{EHAct} is consistent with the Euclidean signature.}:
\begin{align}
f(R)=-\frac{M_{\mathrm{P}}^2}{2}\left(R-2\Lambda\right)\,.\label{EHAct}
\end{align}
It is well-known that the introduction of higher time derivatives leads to additional propagating degrees of freedom \cite{Ostrogradsky:1850fid}.
In addition to the massless graviton, quadratic curvature invariants generically lead to a massive scalar mode, the scalaron, and a ghostlike massive graviton \cite{Stelle:1976gc, Barth:1983hb}.
The appearance of higher-derivative ghosts in the quantum theory is related to Ostrogradski's theorem \cite{Woodard:2006nt}.
Among the higher-derivative models of gravity, $f(R)$ gravity is special, as it only propagates the extra scalar mode and therefore avoids the ghost problem \cite{Woodard:2006nt}.
Note that Ostrogradski's theorem does not apply to $f(R)$ gravity as the proposition of nondegeneracy is violated \cite{Woodard:2006nt}.
Whether a given theory is degenerate or not depends on its highest derivative structure. This structure is encoded in the principal part $\mathbf{D}(\nabla)$ of the fluctuation operator $\mathbf{F}(\nabla)$, defined in \eqref{FlucOp},
\begin{align}
\mathbf{F}(\nabla)=\mathbf{D}(\nabla)+\mathbf{\Pi}(\nabla)\,,
\end{align}
where, as explained in Appendix \ref{AppUFT}, we collect all lower-derivative parts in the operator $\mathbf{\Pi}$.
The degeneracy of $\mathbf{D}$ can have different origins. 
For gauge theories, the operator, derived from the Hessian of the action $S$, is always degenerate.
Therefore, a gauge-fixing procedure is required to break the gauge degeneracy.
But even if the total gauge-fixed operator is nondegenerate, its principal part can still be degenerate. This is exactly the case for $f(R)$ gravity, where
the principal part $\mathbf{D}$ of the fluctuation operator $\mathbf{F}$ arises from the $f_2(\delta_gR)^2$ structure in \eqref{SecVarAct},
\begin{align}
\tensor{D}{_{\mu\nu}^{\rho\sigma}}(\nabla)={}&f_2\tensor{\gamma}{_{\mu\nu,\alpha\beta}}(\tensor{g}{^\alpha^\beta}\,\Delta +\tensor{\nabla}{^\alpha}\nabla^{\beta})(\tensor{g}{^\rho^\sigma}\,\Delta +\tensor{\nabla}{^\rho}\tensor{\nabla}{^\sigma})\,.
\end{align}
Here, $\tensor{\gamma}{_{\mu\nu,\alpha\beta}}$ is the inverse of the bundle metric \eqref{InvBundleMetric}.\footnote{Note that the degeneracy of $\tensor{D}{_{\mu\nu}^{\rho\sigma}}(\nabla)$ is independent of $\tensor{\gamma}{_{\mu\nu,\alpha\beta}}$.}
The corresponding principal symbol is obtained by replacing derivatives $\tensor{\nabla}{_\mu}$ by a constant vector $\mathrm{i}\,\tensor{n}{_\mu}$,
\begin{align}
\tensor{D}{_{\mu\nu}^{\rho\sigma}}(n)
={}&f_2\tensor{\gamma}{_{\mu\nu,\alpha\beta}}\left( \tensor{g}{^\alpha^\beta}n^2-\tensor{n}{^{\alpha}}\tensor{n}{^\beta}\right)\left( \tensor{g}{^\rho^\sigma}n^2-\tensor{n}{^\rho}\tensor{n}{^\sigma}\right)\,, \label{PSfR}
\end{align}
where $n^2=\tensor{g}{^\mu^\nu}\tensor{n}{_\mu}\tensor{n}{_\nu}$.
The dyadic structure of the principal symbol leads to its degeneracy $\det\tensor{D}{_{\mu\nu}^{\rho\sigma}}(n)=0$ and reflects the fact that only the conformal mode propagates with higher-derivatives---not the other components of $\tensor{h}{_\mu_\nu}$.
If we had included other curvature invariants, such as $\tensor*{R}{_\mu_\nu}\tensor*{R}{^\mu^\nu}$ or $\tensor{R}{_\mu_\nu_\rho_\sigma}\tensor{R}{^\mu^\nu^\rho^\sigma}$, in the action \eqref{fRAct}, the principal symbol would, in addition to the structures present in \eqref{PSfR}, contain structures of the form
\begin{align}
\tensor*{\delta}{^{\mu\nu}_{\rho\sigma}}n^4\,,\quad \tensor*{\delta}{^{(\mu}_{(\rho}}\tensor{n}{^{\nu)}}\tensor{n}{_{\sigma)}}n^2\,.
\end{align}
In this case, the transversal-traceless components of $\tensor{h}{_\mu_\nu}$ would propagate also with higher-derivatives. 
In particular, the ``identity'' structure $\tensor*{\delta}{^{\mu\nu}_{\rho\sigma}}n^4$, induced by curvature invariants such as $\tensor{R}{_{\mu\nu}}\tensor{R}{^{\mu\nu}}$, necessary for the principal symbol to be invertible, would lead to the propagation of the additional massive spin-two ghost \cite{Stelle:1976gc}.

The degeneracy of the principal symbol explains why the Ostrogradski instability is avoided in $f(R)$ gravity, but it raises another problem.
In order to obtain the Green's function $\mathbf{G}=-\mathbf{1}/F$, the operator $\mathbf{F}$ needs to be inverted.
In general, an exact inversion of $\mathbf{F}$ is impossible. However, the UV dominant contributions to $\mathbf{G}$ can be obtained in perturbations
\begin{align}
\mathbf{G}=-\frac{\mathbf{1}}{D+\Pi}=-\frac{\mathbf{1}}{D}+\frac{\mathbf{1}}{D}\,\mathbf{\Pi}\,\frac{\mathbf{1}}{D}+\dots
\end{align}
Even if the total operator $\mathbf{F}$ is invertible, this perturbative expansion is not available if $\mathbf{D}$ is degenerate. Therefore, in this respect the situation for $f(R)$ gravity is even more complicated than in fourth-order gravity, where the presence of the $R_{\mu\nu}R^{\mu\nu}$ structure ensures that the principal symbol is nondegenerate and standard methods are applicable \cite{Avramidi:1985ki}.  

In \cite{Barvinsky:1985an}, two methods are proposed to deal with such a degenerate principal symbol.
The first method is based on the inclusion of lower-derivative structures in the definition of the principal part, in order to explicitly break its degeneracy.
But even if the extended principal symbol satisfies the ``generalized causality condition'' \cite{Barvinsky:1985an}, it is not guaranteed that the generalized Schwinger-DeWitt formalism is efficient. 

The second method is the ``method of squaring'' \cite{Barvinsky:1985an}, which is however difficult to realize if components of a relativistic field enter the fluctuation operator with a different number of derivatives.

While the irreducible decomposition of $\tensor{h}{_\mu_\nu}$ generally leads to essential simplification in the derivative structure of the individual components, at the same time the fluctuation operator $\mathbf{F}$ becomes matrix valued. Apart from this complication, the irreducible components are subjected to differential constraints. In particular, this means that functional traces have to be evaluated over constraint fields, which, in curved spacetime, requires the use of complicated nonlocal projection operators; see e.g. \cite{Benedetti:2010nr,Groh:2011dw}. Therefore this method has been almost exclusively applied to highly symmetric backgrounds such as de Sitter space; see e.g. \cite{Cognola:2005de, Codello:2008vh}.

In this paper, we propose a different strategy, which exploits the dyadic structure of the principal part \eqref{PSfR}. In fact, the problems that arise from the degenerate principal symbol in the standard methods lead to a simplification in our approach. It allows us to reduce the calculation to the evaluation of three functional traces---without the aforementioned complications associated with the decomposition of the field $\tensor{h}{_\mu_\nu}$.

\subsection{Relevant operators}\label{SubSec:GaugeFixingFluctuationOp}
\subsubsection{Gauge fixing and ghost operator}
Gauge transformations of the dynamical field $\tensor{h}{_\mu_\nu}$ correspond to diffeomorphisms, infinitesimally generated by the Lie derivative ${\cal L}_{\xi}$ along the vector field $\xi^{\mu}$,
\begin{align}
\delta_{\xi}\tensor{h}{_\mu_\nu}\coloneqq \tensor*{h}{^{\xi}_\mu_\nu}
=\tensor{({\cal L}_{\xi}g)}{_\mu_\nu}
=2\tensor{\xi}{_{(\mu;\nu)}}\,.\label{Diffeoh}
\end{align}
We choose an extension of the de Donder gauge condition
\begin{align}
\chi^{\alpha}[g,h]\coloneqq -\left(\tensor{h}{_\beta^\alpha^{;\beta}}-\frac{1}{2}\tensor{h}{^{;\alpha}}+\tensor{\Upsilon}{^\beta}\tensor{h}{_\beta^\alpha}\right)\label{ExtDeDonder}\,,
\end{align}
which includes a term linear in
\begin{align}
\tensor{\Upsilon}{_{\mu}}\coloneqq\tensor{\left(\ln f_1\right)}{_{;\mu}}=\frac{f_2}{f_1}\,\tensor{R}{_{;\mu}}\,.
\end{align}
The gauge breaking action is given by\footnote{Despite the higher-derivative character of the theory, we have chosen a trivial Nielsen-Kallosh operator $\tensor{O}{_\alpha_\beta}\coloneqq f_1\,\tensor{g}{_\alpha_\beta}$, \cite{Nielsen:1978mp,Kallosh:1978de}. With this choice, \eqref{GBActfR} only affects the lower-derivative part of the fluctuation operator---not its principal part.}
\begin{align}
S_{\rm gb}[g,h]=-\frac{1}{2}\int\mathop{}\!\mathrm{d}^{4}x\, g^{\nicefrac{1}{2}}\left(\tensor{\chi}{^\alpha}\,f_1\,\tensor{g}{_\alpha_\beta}\,\tensor{\chi}{^\beta}\right)\,.\label{GBActfR}
\end{align}
The compensating ghost action reads
\begin{align}
S_{\mathrm{gh}}[g,\overline{\omega},\omega]=\int\mathop{}\!\mathrm{d}^{4}x\, g^{\nicefrac{1}{2}}\,\overline{\omega}_{\alpha}\,\tensor{Q}{^{\alpha}_{\beta}}(\nabla)\,\omega^{\beta}\label{GhostAction}\,,
\end{align}
where the ghost operator $\mathbf{Q}$ derives from \eqref{ExtDeDonder},  
\begin{align}
\tensor{Q}{^{\alpha}_{\beta}}(\nabla)\coloneqq{}&\frac{\delta \chi^{\alpha}[h^{\xi}]}{\delta\xi^{\beta}}\nonumber\\
={}&\tensor*{\delta}{^{\alpha}_{\beta}}\Delta-\tensor{\Upsilon}{_\beta}\nabla^{\alpha}-\tensor*{\delta}{^{\alpha}_{\beta}}\tensor{\Upsilon}{^\mu}\tensor{\nabla}{_\mu}-\tensor*{R}{^{\alpha}_{\beta}}\,.
\label{GhostOP}
\end{align}

\subsubsection{Fluctuation operator}
The fluctuation operator of $f(R)$ gravity ${\mathbf{F}\colon{\cal F}^2\to{\cal F}^2}$, defined by the Hessian of the gauge-fixed action ${S_{\mathrm{tot}}=S+S_{\mathrm{gb}}}$, is a local, nonminimal fourth-order differential operator with components
\begin{align}
\tensor{F}{_{\mu\nu}^{\rho\sigma}}(\nabla)\delta(x,x')\coloneqq  g^{\nicefrac{-1}{2}}\, \tensor{\gamma}{_{\mu\nu,\alpha\beta}}\left.\frac{\delta^2 S_{\mathrm{tot}}}{\delta \tensor{g}{_\alpha_\beta}(x)\delta \tensor{g}{_\rho_\sigma}(x')}\right|_{g=\bar{g}}\,.
\end{align}
It can be represented in the form
\begin{align}
\mathbf{F}=f_1^{\nicefrac{1}{2}}\,\left(\mathbf{A}^{\dagger}\,\frac{f_2}{f_1}\,\mathbf{A}-\mathbf{H}\right)f_{1}^{\nicefrac{1}{2}}\,.\label{DefOpF}
\end{align}
The individual operators, appearing in \eqref{DefOpF}, are explained below. The operator ${\mathbf{A}\colon {\cal F}^2\to{\cal F}^0}$ is defined in terms of the operator ${\mathbf{\bar{A}}\colon {\cal F}^2\to{\cal F}^0}$ by
\begin{align}
\mathbf{A}\coloneqq f_1^{\nicefrac{1}{2}}\mathbf{\bar{A}}f_1^{-\nicefrac{1}{2}}\,.\label{DefOpA}
\end{align}
The operator $\mathbf{\bar{A}}$, in turn, is defined by the first variation of the Ricci scalar $\delta_{g}R=\bar{A}^{\mu\nu}(\nabla)\tensor{h}{_\mu_\nu}$,
\begin{align}
\bar{A}^{\mu\nu}(\nabla)\coloneqq \tensor{g}{^\mu^\nu}\Delta +\nabla^{(\mu}\nabla^{\nu)}-\tensor*{R}{^\mu^\nu}\,.\label{AOP} 
\end{align}
In view of the definition \eqref{DefOpA}, the components of $\mathbf{A}$ involve extra terms compared to the components of $\mathbf{\bar{A}}$,   
\begin{align}
A^{\mu\nu}(\nabla)={}&\tensor{g}{^\mu^\nu}\Delta+\nabla^{(\mu}\nabla^{\nu)}\nonumber\\
&+\tensor{g}{^\mu^\nu}\tensor{\Upsilon}{^\rho}\tensor{\nabla}{_\rho}-\tensor{\Upsilon}{^{(\mu}}\tensor{\nabla}{^{\nu)}}+ \tensor{W}{^\mu^\nu},\\
\tensor{W}{^\mu^\nu}\coloneqq{}& \frac{1}{4}\tensor{\Upsilon}{^\mu}\tensor{\Upsilon}{^\nu}-\frac{1}{2}\tensor{\Upsilon}{^{(\mu;\nu)}}-\frac{1}{4}\tensor{g}{^\mu^\nu}\left(\tensor{\Upsilon}{_\alpha}\tensor{\Upsilon}{^\alpha}\right)\nonumber\\
&+\frac{1}{2}\tensor{g}{^\mu^\nu}\left(\tensor{\Upsilon}{_\alpha^{;\alpha}}\right)-\tensor*{R}{^\mu^\nu}\,.
\end{align}
As explained in \eqref{AdjointInnerProd}, the formal adjoint  ${\mathbf{A}^{\dagger}\colon {\cal F}^0\to{\cal F}^2}$ of $\mathbf{A}$ is defined by
\begin{align}
\langle\mathbf{A}^{\dagger}\,\varphi,h\rangle_2=\langle\varphi,\mathbf{A}\,h\rangle_0\,,\quad  \varphi\in{\cal F}^0,\quad h\in{\cal F}^2\,.
\end{align}
The components of $\mathbf{A}^{\dagger}$ can be expressed in terms of the components of $\mathbf{\bar{A}}$,
\begin{align}
A^{\dagger}_{\rho\sigma}(\nabla)=\tensor{\gamma}{_{\rho\sigma,\mu\nu}}\,f^{-\nicefrac{1}{2}}_{1}\bar{A}^{\mu\nu}(\nabla)f^{\nicefrac{1}{2}}_{1}\,.
\end{align}
The formally self-adjoint, minimal second-order operator ${\mathbf{H}\colon {\cal F}^{2}\to{\cal F}^{2}}$ has components 
\begin{align}
\tensor{H}{_{\alpha\beta}^{\mu\nu}}(\nabla)=\tensor*{\delta}{^{\mu\nu}_{\alpha\beta}}\,\Delta+\tensor{P}{_{\alpha\beta}^{\mu\nu}}\,.\label{DefOpH}
\end{align}
The potential is defined by
\begin{align}
\tensor{P}{_{\gamma\delta}^{\rho\sigma}}\coloneqq{}&-2\tensor{R}{^{\rho}_{(\gamma}^\sigma_{\delta)}}-2\tensor*{\delta}{^{(\rho}_{(\gamma}}\tensor*{R}{^{\sigma)}_{\delta)}}+\tensor{g}{_\gamma_\delta}\tensor*{R}{^\rho^\sigma}\nonumber\\
&+\tensor{g}{^\rho^\sigma}\tensor*{R}{_\gamma_\delta}-\frac{1}{2}\tensor{g}{_\gamma_\delta}\tensor{g}{^\rho^\sigma}R+\frac{f}{f_1}\tensor*{\delta}{^{\rho\sigma}_{\gamma\delta}}\nonumber\\
&
+2\tensor*{\delta}{^{(\rho}_{(\gamma}}\tensor{\Upsilon}{^{\sigma)}_{;\delta)}}
+4\delta^{(\rho}_{(\gamma}\tensor{\Upsilon}{^{\sigma)}}\tensor{\Upsilon}{_{\delta)}}
\nonumber\\
&-\tensor{g}{^\rho^\sigma}\tensor{\Upsilon}{_{\gamma;\delta}}
-\tensor{g}{_\gamma_\delta}\tensor{\Upsilon}{^\rho}\tensor{\Upsilon}{^\sigma}
-\tensor{g}{^\rho^\sigma}\tensor{\Upsilon}{_\gamma}\tensor{\Upsilon}{_\delta}
\nonumber\\
&+\frac{1}{2}\tensor{g}{_\gamma_\delta}\tensor{g}{^\rho^\sigma}\left(\tensor{\Upsilon}{_\alpha^{;\alpha}}+\tensor{\Upsilon}{_\alpha}\tensor{\Upsilon}{^\alpha}\right)\nonumber\\
&-\frac{1}{4}\tensor*{\delta}{^{\rho\sigma}_{\gamma\delta}}\left(6\tensor{\Upsilon}{_\alpha^{;\alpha}}+7\tensor{\Upsilon}{_\alpha}\tensor{\Upsilon}{^\alpha}\right)\,.\label{Ppot}
\end{align}

\section{Calculation of the one-loop divergences}\label{Sec:CalcOneLoopDiv}
The divergent part of the one-loop effective action is the sum of the following traces:
\begin{align}
\left. \varGamma_1 \right|^{\mathrm{div}}=\frac{1}{2}\left. \operatorname{Tr}_{2}\ln\mathbf{F} \right|^{\mathrm{div}}-\left. \operatorname{Tr}_{1}\ln\mathbf{Q} \right|^{\mathrm{div}}\,.\label{Gam1LoopDivFQ}
\end{align}
We rearrange the tensor trace in \eqref{Gam1LoopDivFQ} as follows
\begin{align}
&\left. \operatorname{Tr}_{2}\ln\mathbf{F} \right|^{\mathrm{div}}\nonumber\\
={}&\left. \operatorname{Tr}_{2}\ln\left[ f_1^{\nicefrac{1}{2}}\,\left(\mathbf{A}^{\dagger}\,\frac{f_2}{f_1}\,\mathbf{A}-\mathbf{H}\right)f_{1}^{\nicefrac{1}{2}}\right] \right|^{\mathrm{div}}\nonumber\\
={}&\left. \operatorname{Tr}_{2}\ln\left(\mathbf{A}^{\dagger}\,\frac{f_2}{f_1}\,\mathbf{A}-\mathbf{H}\right) \right|^{\mathrm{div}}\nonumber\\
={}&\left. \operatorname{Tr}_{2}\ln\left(\mathbf{A}^{\dagger}\,\frac{f_2}{f_1}\,\mathbf{A}\,\frac{\mathbf{1}}{H}-\mathbf{1}\right) \right|^{\mathrm{div}}+\left. \operatorname{Tr}_{2}\ln\mathbf{H} \right|^{\mathrm{div}}\,.\label{OPIdFtoH}
\end{align}
In the third equality we have used that for two operators $\mathbf{L}_1$ and $\mathbf{L}_2$, we can formally write\footnote{The failure of this property is usually denoted ``multiplicative anomaly''; see e.g.\ \cite{Kontsevich:1994xe}.}
\begin{align}
\operatorname{Tr}\ln\left(\mathbf{L}_1\mathbf{L}_2\right)=\operatorname{Tr}\ln\mathbf{L}_1+\operatorname{Tr}\ln\mathbf{L}_2\,.\label{MultId}
\end{align} 
For the Einstein-Hilbert case, $f_2=0$, the first trace in \eqref{OPIdFtoH} vanishes and \eqref{Gam1LoopDivFQ} reduces to
\begin{align}
\left. \varGamma_1 \right|^{\mathrm{div}}_{f_2=0}=\frac{1}{2}\left. \operatorname{Tr}_{2}\ln\mathbf{H} \right|^{\mathrm{div}}-\left. \operatorname{Tr}_{1}\ln\mathbf{Q} \right|^{\mathrm{div}}\,.\label{TrLogDivf2Zero}
\end{align}
In this case, the fourth-order operator $\mathbf{F}$ reduces to a second-order operator and the additional scalar degree of freedom is absent.
For $f_2\neq0$, we convert the first trace in \eqref{OPIdFtoH} into a scalar trace,
\begin{align}
&\left. \operatorname{Tr}_{2}\ln\left(\mathbf{A}^{\dagger}\,\frac{f_2}{f_1}\,\mathbf{A}\,\frac{\mathbf{1}}{H}-\mathbf{1}\right) \right|^{\mathrm{div}}=\left. \operatorname{Tr}_{0}\ln\mathbf{S} \right|^{\mathrm{div}}\,,\label{SId}
\end{align}
where $\mathbf{S}$ is the nonlocal scalar operator
\begin{align}
\mathbf{S}\coloneqq{}&\frac{f_1}{f_2}-\mathbf{A}\,\frac{\mathbf{1}}{H}\,\mathbf{A}^{\dagger}\label{ScalOP}\,.
\end{align}
Formally, the identity \eqref{SId} is derived by expanding the logarithm, using the cyclic property of the trace and resumming the terms again.
Note that this manipulation relies on the dyadic nature of the $\mathbf{A}^{\dagger}\mathbf{A}$ structure.
In this way, \eqref{Gam1LoopDivFQ}, which involves the trace of the nonminimal fourth-order operator $\mathbf{F}$, reduces to
\begin{align}
\left. \varGamma_1 \right|^{\mathrm{div}}={}&\frac{1}{2}\left. \operatorname{Tr}_{2}\ln\mathbf{H} \right|^{\mathrm{div}}-\left. \operatorname{Tr}_{1}\ln\mathbf{Q} \right|^{\mathrm{div}}+\frac{1}{2}\left. \operatorname{Tr}_{0}\ln\mathbf{S} \right|^{\mathrm{div}}\,.\label{Gam1Full}
\end{align}
The tensor and vector traces are easily evaluated by standard methods; see Appendix \ref{SubSec:OneLoopDiv}. The evaluation of the scalar trace constitutes the nontrivial part of the calculation as the operator $\mathbf{S}$ involves the inverse of $\mathbf{H}$.
In the following subsections, we separately calculate the tensor trace, the vector trace and the scalar trace in \eqref{Gam1Full}.

\subsection{Divergent part of the tensor trace}\label{SubSubSec:DivTensorTrace}
The tensor trace in \eqref{Gam1Full} is calculated directly by the Schwinger-DeWitt algorithm \eqref{TRLNDiv}, as $\mathbf{H}$ is a minimal second-order operator 
\begin{align}
\mathbf{H}=\Delta+\mathbf{P}\,.
\end{align}
Inserting the potential $\mathbf{P}$, defined in \eqref{Ppot} and the bundle curvature $\tensor{\mathscr{R}}{_{\mu\nu}}$, provided in Table \ref{Table1}, into the general formula \eqref{TRLNDiv}, we obtain 
\begin{align}
&\left. \operatorname{Tr}_2\ln\mathbf{H} \right|^{\mathrm{div}}\nonumber\\
={}&\frac{1}{16\pi^2\varepsilon}\int{\rm d}^4x\, g^{\nicefrac{1}{2}}\left[
-\frac{19}{18}\,\mathcal{G}
-\frac{7}{6}\tensor*{R}{_\mu_\nu}\tensor*{R}{^\mu^\nu}
-5\left(\frac{f}{f_1}\right)^2\right.\nonumber\\
&+\frac{17}{3}\frac{f}{f_1}R
-\frac{5}{4}R^2
+9\frac{f}{f_1}\left(\tensor{\Upsilon}{_\mu^{;\mu}}\right)
-\frac{9}{2}R\left(\tensor{\Upsilon}{_\mu^{;\mu}}\right)\nonumber\\
&-\frac{17}{4}R\left(\tensor{\Upsilon}{_\mu}\tensor{\Upsilon}{^\mu}\right)
+\frac{15}{2}\frac{f}{f_1}\left(\tensor{\Upsilon}{_\mu}\tensor{\Upsilon}{^\mu}\right)
+\tensor{R}{_\mu_\nu}\tensor{\Upsilon}{^\mu}\tensor{\Upsilon}{^\nu}\nonumber\\
&\left.-\frac{21}{4}\left(\tensor{\Upsilon}{_\mu^{;\mu}}\right)^2-\frac{3}{4}\left(\tensor{\Upsilon}{_\mu}\tensor{\Upsilon}{^\mu}\right)\left(\tensor{\Upsilon}{_\nu^{;\nu}}\right)
-\frac{141}{16}\left(\tensor{\Upsilon}{_\mu}\tensor{\Upsilon}{^\mu}\right)^2
\right]\,.\label{TrLogDivTensor}
\end{align}
The Gauss-Bonnet term is defined as
\begin{align}
\mathcal{G}\coloneqq{}&\tensor{R}{_{\mu\nu\rho\sigma}}\tensor{R}{^{\mu\nu\rho\sigma}}-4\tensor*{R}{_{\mu\nu}}\tensor*{R}{^{\mu\nu}}+R^2\,.\label{GaussBonnet}
\end{align}

\subsection{Divergent part of the vector trace}\label{SubSubSec:DivVectorTrace}
In contrast to $\mathbf{H}$, the ghost operator $\mathbf{Q}$ is not yet of the form \eqref{MinSecOp}, suitable for a direct application of the Schwinger-DeWitt algorithm \eqref{TRLNDiv}, as it contains terms linear in derivatives. We write \eqref{GhostOP} as
\begin{align}
\mathbf{Q}=\Delta-2\tensor{\mathbf{\Omega}}{^\mu}\tensor{\nabla}{_\mu}+\mathbf{U}\,,\label{GhOp2}
\end{align}
where the coefficients of $\mathbf{\Omega}^{\mu}$ and $\mathbf{U}$ are given by
\begin{align}
\tensor{{[\tensor{\Omega}{^\mu}]}}{^{\alpha}_{\beta}}\coloneqq\frac{1}{2}\left(\tensor*{\delta}{^{\alpha}_{\beta}}\tensor{\Upsilon}{^{\mu}}+\tensor{g}{^\mu^\alpha}\tensor{\Upsilon}{_\beta}\right)\,,\quad
\tensor{U}{^{\alpha}_{\beta}}\coloneqq-\tensor*{R}{^{\alpha}_{\beta}}\,.
\end{align}
By redefining the covariant derivative
$\tensor*{\nabla}{^\prime_\mu}=\tensor{\nabla}{_\mu}+\mathbf{\Omega}_\mu$,
the operator is brought into standard form,
\begin{align}
\mathbf{Q}=\Delta^\prime+\mathbf{U}^\prime\,,\label{GhOp3}
\end{align}
at the price of a modified potential and bundle curvature:
\begin{align}
\tensor{{U^\prime}}{^{\alpha}_{\beta}}={}&-\tensor{R}{^{\alpha}_{\beta}}+\tensor{{[\tensor{\Omega}{^\mu}]}}{^{\alpha}_{\beta;\mu}}+ \tensor{{[\tensor{\Omega}{^\mu}]}}{^{\alpha}_{\gamma}} \tensor{{[\tensor{\Omega}{^\nu}]}}{^{\gamma}_{\beta}}\label{ModPGhost}\,\tensor{g}{_\mu_\nu}\,,\\
\tensor{{[\tensor*{\mathscr{R}}{^\prime_\mu_\nu}]}}{^{\alpha}_{\beta}}={}& \tensor{{[\tensor{\mathscr{R}}{_\mu_\nu}]}}{^{\alpha}_{\beta}}+\tensor{{[\tensor{\Omega}{^\rho}]}}{^{\alpha}_{\beta;[\mu}}\tensor{g}{_{\nu]}_\rho}\nonumber\\
&+\tensor{{[\tensor{\Omega}{^\rho}]}}{^{\alpha}_{\gamma}}\tensor{{[\tensor{\Omega}{^\sigma}]}}{^{\gamma}_{\beta}}\tensor{g}{_\rho_{[\mu}}\tensor{g}{_{\nu]}_\sigma}\,.\label{ModCurl}
\end{align}
Inserting the minimal second-order operator \eqref{GhOp3} together with \eqref{ModCurl} into the general formula \eqref{TRLNDiv}, we obtain the divergent contribution of the ghost trace:
\begin{align}
&\left. \operatorname{Tr}_1\ln\mathbf{Q} \right|^{\mathrm{div}}\nonumber\\
={}&\frac{1}{16\pi^2\varepsilon}\int{\rm d}^4x\, g^{\nicefrac{1}{2}}\left[
\frac{11}{180}\mathcal{G}
-\frac{7}{30}\tensor*{R}{_\mu_\nu}\tensor*{R}{^\mu^\nu}
-\frac{17}{60}R^2\right.\nonumber\\
&+\frac{13}{12}R\left(\tensor{\Upsilon}{_\mu^{;\mu}}\right)
+\frac{13}{24}R\left(\tensor{\Upsilon}{_\mu}\tensor{\Upsilon}{^\mu}\right)
+\tensor{R}{_\mu_\nu}\tensor{\Upsilon}{^\mu}\tensor{\Upsilon}{^\nu}\nonumber\\
&\left.-\frac{7}{8}\left(\tensor{\Upsilon}{_\mu^{;\mu}}\right)^2-\frac{7}{8}\left(\tensor{\Upsilon}{_\mu}\tensor{\Upsilon}{^\mu}\right)\left(\tensor{\Upsilon}{_\nu^{;\nu}}\right)
-\frac{19}{32}\left(\tensor{\Upsilon}{_\mu}\tensor{\Upsilon}{^\mu}\right)^2
\right]\,.
\label{TrlogDivVec}
\end{align}

\subsection{Divergent part of the scalar trace}\label{SubSubSec:DivScalarrTrace}
It remains to calculate the scalar trace in \eqref{Gam1Full}.
The operator $\mathbf{S}$ involves a nonlocal part due to the inverse of the operator $\mathbf{H}$. Therefore, the functional trace cannot be evaluated directly with \eqref{TRLNDiv}. Nevertheless, $\mathbf{S}$ is a scalar operator, 
\begin{align}
\mathbf{S}=3\,\Delta+{\cal O}\left(\mathfrak{M}\right)\,,\label{Sop}
\end{align}
with principal part $3\Delta$ and lower-derivative terms ${\cal O}(\mathfrak{M})$, which we treat as perturbations. For the  divergent part in four dimensions, it is sufficient to expand up to ${\cal O}\left(\mathfrak{M}^4\right)$. In order to determine the terms ${\cal O}(\mathfrak{M})$ in \eqref{Sop} explicitly, we make use of the following operator identity:
\begin{align}
\mathbf{A}\frac{\mathbf{1}}{H}={}&\frac{1}{K}\,\mathbf{A}+\frac{1}{K}\mathbf{B}\frac{\mathbf{1}}{H}\,.\label{OpIds}
\end{align}
The minimal operator ${\mathbf{K}\,\colon{\cal F}^0\to{\cal F}^{0}}$ is defined as
\begin{align}
\mathbf{K}\coloneqq\Delta-R+\frac{f}{f_1}\,.
\end{align}
The operator ${\mathbf{B}\colon {\cal F}^2\to{\cal F}^0}$  and its adjoint ${\mathbf{B}^{\dagger}\colon {\cal F}^2\to{\cal F}^0}$ are second-order operators, which are ${\cal O}\left(\mathfrak{M}^2\right)$.
This property is crucial for the efficient use of the operator identity \eqref{OpIds}. In components, $\mathbf{B}$ and $\mathbf{B}^{\dagger}$ read 
\begin{align}
\tensor{B}{^\mu^\nu}(\nabla)={}&\tensor*{B}{_2^\mu^\nu^\rho^\sigma}\tensor{\nabla}{_\rho}\tensor{\nabla}{_\sigma}+\tensor*{B}{_3^\mu^\nu^\rho}\tensor{\nabla}{_\rho}+\tensor*{B}{_4^\mu^\nu},\label{BCoeff}\\
\tensor*{B}{^\dagger_\mu_\nu}(\nabla)={}&\tensor{\gamma}{_{\mu\nu,\alpha\beta}}\left(\tensor{\nabla}{_\rho}\tensor{\nabla}{_\sigma}\tensor{B}{_2^{\alpha\beta\rho\sigma}}-\tensor{\nabla}{_\rho}\tensor*{B}{_3^{\alpha\beta\rho}}+\tensor*{B}{_4^{\alpha\beta}}\right)\,,
\end{align}
where the coefficients $B_{i}={\cal O}(\mathfrak{M}^i)$ are local background tensors and explicitly presented in \eqref{AppCoefB2}--\eqref{AppCoefB4}.
We use the identity \eqref{OpIds} and its adjoint to write
\begin{align}
\left. \operatorname{Tr}_0\ln\mathbf{S} \right|^{\mathrm{div}}
={}&\operatorname{Tr}_0\ln\left[\frac{1}{K}\left(-\mathbf{A}\mathbf{A}^{\dagger}\mathbf{K}-\mathbf{B}\mathbf{A}^{\dagger}\phantom{\frac{1}{\mathbf{H}}}\right.\right.\nonumber\\
&\left.\left.\left. +\mathbf{K}\frac{f_1}{f_2}\mathbf{K}-\mathbf{B}\frac{\mathbf{1}}{H}\mathbf{B}^{\dagger}\right)\frac{1}{K}\right] \right|^{\mathrm{div}}\,.\label{TrLogDivScalOpId}
\end{align}
Next, we define the sixth-order operator ${\mathbf{X}\colon{\cal F}^0\to{\cal F}^0}$,
\begin{align}
\mathbf{X}\coloneqq\frac{1}{3}\left(-\mathbf{A}\mathbf{A}^{\dagger}\mathbf{K}-\mathbf{B}\mathbf{A}^{\dagger}+\mathbf{K}\frac{f_1}{f_2}\mathbf{K}\right)\,.
\end{align}
Using the property \eqref{MultId}, we write \eqref{TrLogDivScalOpId} in the compact form
\begin{align}
\left. \operatorname{Tr}_0\ln\mathbf{S} \right|^{\mathrm{div}}={}&\left. \operatorname{Tr}_0\ln\left(\mathbf{X}-\frac{1}{3}\mathbf{B}\frac{\mathbf{1}}{H}\mathbf{B}^{\dagger}\right) \right|^{\mathrm{div}}\nonumber\\
&-2\left. \operatorname{Tr}_0\ln \mathbf{K} \right|^{\mathrm{div}}\,.
\label{TrLogDivX}
\end{align}
In components, the operator $\mathbf{X}$ reads
\begin{align}
X(\nabla)={}&\Delta^3+\tensor*{X}{_2^\mu^\nu}\tensor{\nabla}{_\mu}\tensor{\nabla}{_\nu}\,\Delta+\tensor*{X}{_3^\mu^\nu^\rho}\tensor{\nabla}{_\mu}\tensor{\nabla}{_\nu}\tensor{\nabla}{_\rho}\nonumber\\
&+\tensor*{X}{_4^\mu^\nu}\tensor{\nabla}{_\mu}\tensor{\nabla}{_\nu}+{\cal O}(\mathfrak{M}^5)\,.\label{OpX}
\end{align}
The coefficients $X_i={\cal O}\left(\mathfrak{M}^i\right)$ are totally symmetric local background tensors. The formal self-adjointness of $\mathbf{X}$ leads to essential simplifications: the term linear in the background dimension is absent, such that the perturbative expansion starts with $\tensor*{X}{_2^{\mu\nu}}={\cal O}(\mathfrak{M}^2)$. Moreover, $\tensor*{X}{_2^{\mu\nu}}$ only has two free indices instead of four, as two derivatives in \eqref{OpX} are contracted into a Laplacian $\Delta$.
The explicit coefficients $X_i$ can be found in \eqref{AppCoefX2}--\eqref{AppCoefX4}. 

In order to extract the divergent part of the trace \eqref{TrLogDivX}, we first treat the nonlocal term as perturbation and expand the logarithm up to terms ${\cal O}(\mathfrak{M}^4)$,
\begin{align}
&\left. \operatorname{Tr}_0\ln\left(\mathbf{X}-\frac{1}{3}\mathbf{B}\frac{\mathbf{1}}{H}\mathbf{B}^{\dagger}\right) \right|^{\mathrm{div}}\nonumber\\
=&{}\left. \operatorname{Tr}_0\ln\mathbf{X} \right|^{\mathrm{div}}-\frac{1}{3}\left. \operatorname{Tr}_0\left(\frac{1}{X}\mathbf{B}\frac{\mathbf{1}}{H}\mathbf{B}^{\dagger}\right) \right|^{\mathrm{div}}\,.\label{ScalTrace2}
\end{align}
Since the second trace in \eqref{ScalTrace2} is already ${\cal O}\left(\mathfrak{M}^4\right)$, we freely commute all operators, use
\begin{align}
\frac{\mathbf{1}}{H}=\mathbf{1}\,\frac{1}{\Delta}+{\cal O}(\mathfrak{M})\,,\quad \frac{1}{X}=\frac{1}{\Delta^3}+{\cal O}(\mathfrak{M})\,,
\end{align}
and reduce \eqref{ScalTrace2} to the following functional trace:
\begin{align}
&\left. \operatorname{Tr}_0\left(\frac{1}{X}\mathbf{B}\frac{\mathbf{1}}{H}\mathbf{B}^{\dagger}\right) \right|^{\mathrm{div}}=\left. \operatorname{Tr}_0\left(\mathbf{B}\mathbf{B}^{\dagger}\frac{1}{\Delta^4}\right) \right|^{\mathrm{div}}\nonumber\\
={}&\int\mathrm{d}^4x\,\tensor*{B}{_2^\mu^\nu^\rho^\sigma}\tensor{\gamma}{_{\mu\nu,\alpha\beta}}\tensor*{B}{_2^\alpha^\beta^\gamma^\delta}\mathscr{U}^{(4,4)}_{\rho\sigma\gamma\delta}\,.
\end{align}
Here, $\mathscr{U}^{(4,4)}_{\rho\sigma\gamma\delta}$ denotes a universal functional trace, defined in Appendix \ref{AppUFT}.
Next, we extract the divergent part from the first trace in \eqref{ScalTrace2}. We insert the representation \eqref{OpX} for $\mathbf{X}$, expand the logarithm around $\Delta^3$ up to  ${\cal O}(\mathfrak{M}^4)$ and obtain again a sum of universal functional traces,
\begin{align}
&\left. \operatorname{Tr}_0\ln \mathbf{X} \right|^{\mathrm{div}}= 3\left. \operatorname{Tr}_0\ln \Delta \right|^{\mathrm{div}}+\int\mathrm{d}^4x\,\tensor*{X}{_2^\mu^\nu}\mathscr{U}^{(2,2)}_{\mu\nu}\nonumber\\
&+\int\mathrm{d}^4x\,\tensor*{X}{_3^\mu^\nu^\rho}\mathscr{U}^{(3,3)}_{\mu\nu\rho}+\int\mathrm{d}^4x\,\tensor*{X}{_4^\mu^\nu}\mathscr{U}^{(2,3)}_{\mu\nu}\nonumber\\
&-\frac{1}{2}\int\mathrm{d}^4x\,\tensor*{X}{_2^\mu^\nu}\tensor*{X}{_2^\rho^\sigma}\mathscr{U}^{(4,4)}_{\mu\nu\rho\sigma}\,.
\end{align} 
The trace including $\mathbf{K}$ in \eqref{TrLogDivX} is evaluated directly with \eqref{TRLNDiv}.
Inserting the explicit expressions for $B_2^{\mu\nu\rho\sigma}$, $X^{\mu\nu}_2$, $X^{\mu\nu\rho}_3$ and $X_4^{\mu\nu}$, tabulated in \eqref{AppCoefB2}, \eqref{AppCoefX2}--\eqref{AppCoefX4}, together with the corresponding universal functional traces $\mathscr{U}^{(p,n)}_{\mu_1\cdots\mu_p}$, tabulated in \eqref{UFT1}--\eqref{UFTs}, we find for the divergent part of the scalar trace \eqref{TrLogDivX},
\begin{widetext}
\begin{align}
\left.\operatorname{Tr}_0\ln\mathbf{S}\right|^{\mathrm{div}}={}\frac{1}{16\pi^2\varepsilon}\int{\rm d}^4x\, g^{\nicefrac{1}{2}}&\left[
-\frac{1}{180}\mathcal{G}
-\frac{133}{180}\tensor*{R}{_\mu_\nu}\tensor*{R}{^\mu^\nu}
+\frac{1}{2}\left(\frac{f}{f_1}\right)^2
-\frac{1}{18}\left(\frac{f_1}{f_2}\right)^2
-\frac{7}{6}\frac{f}{f_1}R
+\frac{1}{3}\frac{f}{f_2}
+\frac{187}{360}R^2
\right.\nonumber\\
&\;\,\left.
+\frac{3}{2}\frac{f}{f_1}\left(\tensor{\Upsilon}{_\mu^{;\mu}}\right)
-\frac{19}{12}R\left(\tensor{\Upsilon}{_\mu^{;\mu}}\right)
+\frac{13}{72}R\left(\tensor{\Upsilon}{_\mu}\tensor{\Upsilon}{^\mu}\right)
-\frac{3}{4}\frac{f}{f_1}\left(\tensor{\Upsilon}{_\mu}\tensor{\Upsilon}{^\mu}\right)
+\frac{29}{9}\tensor{R}{_\mu_\nu}\tensor{\Upsilon}{^\mu}\tensor{\Upsilon}{^\nu}
\right.\nonumber\\
&\;\,\left.
-\frac{53}{24}\left(\tensor{\Upsilon}{_\mu^{;\mu}}\right)^2
-\frac{17}{8}\left(\tensor{\Upsilon}{_\mu}\tensor{\Upsilon}{^\mu}\right)\left(\tensor{\Upsilon}{_\nu^{;\nu}}\right)
-\frac{37}{96}\left(\tensor{\Upsilon}{_\mu}\tensor{\Upsilon}{^\mu}\right)^2
\right]\,.\label{TrLogDivScal}
\end{align}
\end{widetext}

\section{One-loop divergences on arbitrary backgrounds -- final result}\label{Sec:OneLoopDivOnGenBackFinalResult}
According to \eqref{Gam1Full}, we add the partial results for the tensor trace \eqref{TrLogDivTensor}, the vector trace \eqref{TrlogDivVec} and the scalar trace \eqref{TrLogDivScal} to obtain the final result for the divergent part of the one-loop divergences for $f(R)$ gravity on an arbitrary background.

Note that all invariants must be homogeneous functions of degree zero under simultaneous rescaling of the function $f$ and its derivatives.
Below, we present the final result in terms of curvature and \mbox{$\Upsilon$-structures}:
\begin{widetext}
\begin{align}
\left. \varGamma_1 \right|^{\mathrm{div}}={}\frac{1}{32\pi^2\varepsilon}\int{\rm d}^4x\, g^{\nicefrac{1}{2}}&\left[
-\frac{71}{60}\mathcal{G}
-\frac{259}{180}\tensor*{R}{_\mu_\nu}\tensor*{R}{^\mu^\nu}
-\frac{9}{2}\left(\frac{f}{f_1}\right)^2
-\frac{1}{18}\left(\frac{f_1}{f_2}\right)^2
+\frac{9}{2}\frac{f}{f_1}R
+\frac{1}{3}\frac{f}{f_2}
-\frac{59}{360}R^2
\right.\nonumber\\
&\;\,\left.
+\frac{21}{2}\frac{f}{f_1}\left(\tensor{\Upsilon}{_\mu^{;\mu}}\right)
-\frac{33}{4}R\left(\tensor{\Upsilon}{_\mu^{;\mu}}\right)
-\frac{371}{72}R\left(\tensor{\Upsilon}{_\mu}\tensor{\Upsilon}{^\mu}\right)
+\frac{27}{4}\frac{f}{f_1}\left(\tensor{\Upsilon}{_\mu}\tensor{\Upsilon}{^\mu}\right)
+\frac{20}{9}\tensor{R}{_\mu_\nu}\tensor{\Upsilon}{^\mu}\tensor{\Upsilon}{^\nu}\right.\nonumber\\
&\;\,\left.
-\frac{137}{24}\left(\tensor{\Upsilon}{_\mu^{;\mu}}\right)^2
-\frac{9}{8}\left(\tensor{\Upsilon}{_\mu}\tensor{\Upsilon}{^\mu}\right)\left(\tensor{\Upsilon}{_\nu^{;\nu}}\right)
-\frac{769}{96}\left(\tensor{\Upsilon}{_\mu}\tensor{\Upsilon}{^\mu}\right)^2
\right]\,.\label{Gamma1LoopOffShellFinalGamPara}
\end{align}
\end{widetext}
This constitutes our main result. The result expressed in terms of different invariants, which are better suited for the reduction to the on shell divergences and the reduction to spaces of constant curvature, are presented in Appendix~\ref{App:RepFinRes}. As expected on general grounds, the one-loop divergences contain up to four derivatives of the Ricci scalar. The presence of the curvature squared structures proportional to $\mathcal{G}$ and $\tensor{R}{_{\mu\nu}}\tensor{R}{^{\mu\nu}}$ and the derivative structures in the second and third line just confirm explicitly that $f(R)$ gravity is perturbatively nonrenormalizable. 
In the form \eqref{Gamma1LoopOffShellFinalEPara}, the on shell reduction is trivially performed by setting $\tensor{E}{_\mu_\nu}=0$,
\begin{align}
\left. \varGamma_1 \right|^{\mathrm{div}}_{\mathscr{E}=0}={}&\frac{1}{32\pi^2\varepsilon}\int{\rm d}^4x\, g^{\nicefrac{1}{2}}\left[
-\frac{71}{60}\mathcal{G}
-\frac{609}{80}\tensor*{R}{_\mu_\nu}\tensor*{R}{^\mu^\nu}\right.\nonumber\\
&+\frac{1}{3}\frac{f}{f_2}
-\frac{115}{288}\left(\frac{f}{f_1}\right)^2
-\frac{1}{18}\left(\frac{f_1}{f_2}\right)^2\nonumber\\
&\left.
-\frac{15}{64}\frac{f}{f_1}\,R
+\frac{3919}{1440}R^2
+\frac{15}{64}R\,\Delta\ln f_1\right]\,.\label{OnShellFinalResult}
\end{align}
Note that $R\,\Delta\ln f_1$ is the only derivative structure that survives the on shell reduction.


\section{$f(R)$ gravity on a de Sitter background}\label{Sec:fRdeSitter} 
The Riemann curvature tensor of a maximally symmetric space is given in terms of the constant scalar curvature $R_0$,
\begin{align}
\tensor{R}{_{\mu\nu\rho\sigma}}={}&\frac{R_0}{12}\left(\tensor*{g}{^{(0)}_\mu_\rho}\tensor*{g}{^{(0)}_\nu_\sigma}-\tensor*{g}{^{(0)}_\mu_\sigma}\tensor*{g}{^{(0)}_\nu_\rho}\right)\,.\label{ConstCurvRiem}
\end{align}
In particular, we have $\tensor{\Upsilon}{_\mu}=0$.
Euclidean de Sitter space in four dimensions is a sphere $S_4$ of constant radius~$r_0$,
\begin{align}
r_0^2=\frac{12}{R_0}\,,\quad R_0>0\,.
\end{align}
The volume $V({S_4})$ is given by
\begin{align}
V({S_4})=\int\mathop{}\!\mathrm{d}^{4}x\,g_{(0)}^{\nicefrac{1}{2}}=\frac{384\,\pi^2}{R_0^2}\,.\label{VolSphered}
\end{align} 

\subsection{One-loop divergences: Reduction to de Sitter space}
For the off-shell one-loop divergences in the basis \eqref{Gamma1LoopOffShellFinalRPara}, the reduction to a constant curvature background becomes trivial:
\begin{align}
\left.\varGamma_1\right|^{\mathrm{div}}_{R_0}=\frac{1}{\varepsilon}&\left[-\frac{173}{20}-54\left(\frac{f}{R_0\,f_1}\right)^2+54\frac{f}{R_0\,f_1}\right.\nonumber\\
&\;\,\left.-\frac{2}{3}\left(\frac{f_1}{R_0\,f_2}\right)^2+4\frac{f}{R_0^2f_2}\right]\,.\label{Gam1LoopdeSitter}
\end{align}
The subsequent on shell reduction is performed by noting that on spaces of constant curvature, the on shell relation \eqref{OnShellScalar} reduces to the algebraic equation
\begin{align}
\mathscr{E}_0\coloneqq-R_0f_1+2f\,.\label{OnShellCondConstCurv}
\end{align} 
Inserting \eqref{OnShellCondConstCurv} into \eqref{Gam1LoopdeSitter} we obtain the on shell one-loop divergences on a de Sitter background:
\begin{align}
\left. \varGamma_1 \right|^{\mathrm{div}}_{R_0,\mathscr{E}_0=0}
={}&\frac{1}{\varepsilon}\left[\frac{97}{20}+4\frac{f}{R_0^2f_2}-\frac{8}{3}\left(\frac{f}{R_0^2f_2}\right)^2\right]
\,.
\label{Gam1LoopOnShelldeSitter}
\end{align}

\subsection{One-loop divergences: Direct calculation}
Beside the reduction of the one-loop divergences to de Sitter space, it is instructive to repeat the calculation directly in de Sitter space.
On a de Sitter background, the operators  \eqref{DefOpH}, \eqref{GhostOP}, \eqref{DefOpA} and  \eqref{ScalOP} reduce to
\begin{align}
\tensor{H}{_{(0)\,\alpha\beta}^{\mu\nu}}&=\tensor*{\delta}{^{\mu\nu}_{\alpha\beta}}\,\Delta+\tensor{P}{_{(0)\,\alpha\beta}^{\mu\nu}}\,,\label{HOperatordeSitter}\\ 
\tensor{Q}{_{(0)\,\beta}^{\alpha}}&=\tensor*{\delta}{^{\alpha}_{\beta}}\,\Delta-\frac{1}{4}R_0\,\tensor*{\delta}{^{\alpha}_{\beta}}\,,\label{QOperatordeSitter}\\
\tensor*{A}{_{(0)}^{\mu\nu}}&=\tensor*{g}{^\mu^\nu_{(0)}}\,\Delta+\nabla^{(\mu}\nabla^{\nu)}-\frac{1}{4}R_0\,\tensor*{g}{^\mu^\nu_{(0)}}\,,\label{AOperatordeSitter}\\
\mathbf{S}_0&=\frac{f_1}{f_2}-\mathbf{A}_0\,\frac{\mathbf{1}}{H_0}\,\mathbf{A}_0^{\dagger}\,,
\end{align}
where $\tensor{P}{_{(0)\,\alpha\beta}^{\mu\nu}}$ in \eqref{Ppot} reduces to the constant potential  
\begin{align}
\tensor{P}{_{(0)\,\alpha\beta}^{\mu\nu}}={}&-\frac{1}{3}R_0\left(\tensor*{\delta}{^{\mu\nu}_{\alpha\beta}}+\frac{1}{2}\tensor*{g}{_\alpha_\beta^{(0)}}\tensor*{g}{^\mu^\nu_{(0)}}\right)+\frac{f}{f_1}\,\tensor*{\delta}{^{\mu\nu}_{\alpha\beta}}\,.
\end{align}
In view of the simple minimal second-order operators $\mathbf{H}_0$ and $\mathbf{Q}_0$, the calculation of the divergent parts of the tensor trace \eqref{TrLogDivTensor} and the vector trace \eqref{TrlogDivVec} is calculated directly with \eqref{TRLNDiv},
\begin{align}
\left. \operatorname{Tr}_{2}\ln\mathbf{H}_0 \right|^{\mathrm{div}}={}&\frac{1}{\varepsilon}\left[- \frac{371}{9}+136\frac{f}{f_1R_0}-120\left(\frac{f}{f_1R_0}\right)^2\right]\,,\label{TTrdeSitter}\\
\left. \operatorname{Tr}_{1}\ln\mathbf{Q}_0 \right|^{\mathrm{div}}={}&-\frac{1}{\varepsilon}\frac{358}{45}\label{VTrdeSitter}\,.
\end{align}
In particular, the operator identity \eqref{OpIds} essentially simplifies, as $\mathbf{B}_0=0$ on a space of constant curvature
\begin{align}
\mathbf{A}_0\,\frac{\mathbf{1}}{H_0}=\frac{1}{K_0}\,\mathbf{A}_0,\quad\quad \mathbf{K}_0=\Delta-R_0+\frac{f}{f_1}\,. \label{AGId}
\end{align}
Therefore, by using \eqref{AGId}, the evaluation of the scalar trace becomes very simple
\begin{align}
\operatorname{Tr}_0 \ln\mathbf{S}_0
={}&\operatorname{Tr}_0\left(\frac{f_1}{f_2}-\frac{1}{K_0}\,\mathbf{A}_0\,\mathbf{A}_0^{\dagger}\right)\,.\label{TrDivScalConstCurv}
\end{align}
The product $\mathbf{A}_0{\mathbf{A}^{\dagger}}_0$ factorizes into two minimal second-order operators,
\begin{align}
\mathbf{A}_0\mathbf{A}_0^{\dagger}=-3\left(\Delta-\frac{1}{3}R_0\right)\left(\Delta-\frac{1}{2}R_0\right)\label{ACAId}\,.
\end{align}
Combining \eqref{TrDivScalConstCurv} with \eqref{ACAId}, we find 
\begin{widetext}
\begin{align}
\left. \operatorname{Tr}_0\ln\mathbf{S}_0 \right|^{\mathrm{div}}
={}&\left. \operatorname{Tr}_0\ln\left[3\left(\Delta-\frac{1}{3}R_0\right)\left(\Delta-\frac{1}{2}R_0\right)+\frac{f_1}{f_2}\left(\Delta-R_0+\frac{f}{f_1}\right)\right] \right|^{\mathrm{div}}-\left. \operatorname{Tr}_0\ln \left(\Delta-R_0+\frac{f}{f_1}\right) \right|^{\mathrm{div}}\nonumber\\
={}&\frac{1}{\varepsilon}\left[\frac{721}{90}-28\frac{f}{f_1R_0}+12\left(\frac{f}{f_1R_0}\right)^2+8\frac{f}{f_2R_0^2}-\frac{4}{3}\left(\frac{f_1}{f_2R_0}\right)^2\right]\,.\label{TrlogDivScalConstCurvOff}
\end{align}
\end{widetext}
The off-shell divergences of the traces in \eqref{TrlogDivScalConstCurvOff} are extracted by the generalized Schwinger-DeWitt formalism \cite{Barvinsky:1985an}.

On shell, the fourth-order operator in the first trace of \eqref{TrlogDivScalConstCurvOff} factorizes into two second-order operators, one of which cancels the contribution from the second trace 
\begin{align}
\left. \operatorname{Tr}_0\ln\mathbf{S}_0 \right|^{\mathrm{div}}_{\mathscr{E}_0=0}=\left. \operatorname{Tr}_0\ln\left(\Delta-\frac{1}{3}R_0+\frac{2}{3}\frac{f}{f_2R_0}\right) \right|^{\mathrm{div}}\,.\label{TrLnDivScaldeSitterOnShell}
\end{align}
Combining \eqref{TTrdeSitter}, \eqref{VTrdeSitter} and \eqref{TrlogDivScalConstCurvOff}, the one-loop divergences directly calculated on a de Sitter background read
\begin{align}
\left. \varGamma_1 \right|^{\mathrm{div}}_{R_0}={}&\frac{1}{2}\left. \operatorname{Tr}_2\ln\,\mathbf{H}_0 \right|^{\mathrm{div}}-\left. \operatorname{Tr}_1\ln\,\mathbf{Q}_0 \right|^{\mathrm{div}}\nonumber\\
&+\frac{1}{2}\left. \operatorname{Tr}_0\ln\mathbf{S}_0 \right|^{\mathrm{div}}\,.
\end{align}
This agrees with the off-shell one-loop divergences \eqref{Gam1LoopdeSitter}, which were obtained by reducing the result for an arbitrary background \eqref{Gamma1LoopOffShellFinalRPara}.

\subsection{One-loop divergences: Irreducible decomposition}\label{SubSec:IrredDec}
An independent calculation for the off-shell divergences is obtained by making use of the decomposition of $\tensor{h}{_\mu_\nu}$ into its irreducible components:
\begin{align}
\tensor{h}{_\mu_\nu}={}\tensor*{h}{^{\perp}_\mu_\nu}
&+\frac{\tensor{g}{_\mu_\nu}}{4}h
+2\tensor{\nabla}{_{(\mu}}\tensor*{v}{^{\perp}_{\nu)}}+2\left(\tensor{\nabla}{_{\mu}}\!\tensor{\nabla}{_{\nu}}b+\frac{\tensor{g}{_\mu_\nu}}{4}\Delta b\right)\,.\label{IrredDec}
\end{align}
Here, $\tensor*{h}{^{\perp}_\mu_\nu}$ is a symmetric transverse-traceless tensor, ${h=\tensor{g}{^\mu^\nu}\tensor{h}{_\mu_\nu}}$ is the trace, $\tensor*{v}{^{\perp}_{\mu}}$ is a transversal vector and $b$ the longitudinal scalar, which are subjected to the differential constraints
\begin{align}
\tensor{g}{^\mu^\nu}\tensor*{h}{^{\perp}_\mu_\nu}=0\,,\quad \tensor{\nabla}{^{\mu}}\tensor*{h}{^{\perp}_\mu_\nu}=0\,,\quad
\tensor{\nabla}{^{\mu}}\tensor*{v}{^{\perp}_{\mu}}=0\,.\label{DiffConstraint}
\end{align}
This decomposition is particularly useful on a de Sitter background for two reasons. First, the fluctuation operator acquires a simple, almost diagonal, block form.
Second, the projection operators, required in the functional traces over the invariant subspaces, are significantly less complicated than on general backgrounds.

Under an infinitesimal diffeomorphism \eqref{Diffeoh}, generated by the vector $\xi_{\mu}=\xi^{\perp}_{\mu}+\tensor{\nabla}{_{\mu}}\xi$ with $\tensor{\nabla}{^\mu}\xi^{\perp}_{\mu}=0$, the field $\tensor{h}{_\mu_\nu}$ transforms as
\begin{align}
\delta_{\xi}\tensor{h}{_\mu_\nu}=2\nabla_{(\mu}\xi_{\nu)}^{\perp}+2\tensor{\nabla}{_\mu}\tensor{\nabla}{_\nu}\xi\,.
\end{align}
Therefore, the individual components change as
\begin{align}
\delta_{\xi}h^{\perp}_{\mu\nu}=0\,,\;\;\delta_{\xi}v^{\perp}_{\mu}=\xi^{\perp}_{\mu}\,,\;\;\delta_{\xi}h=-2\Delta \xi\,,\;\; \delta_{\xi}b=\xi\,.
\end{align}
Clearly, $h^{\perp}_{\mu\nu}$ is gauge invariant, while the transverse vector $v^{\perp}_{\mu}$ and the trace $h$ as well as the longitudinal scalar $b$ are not.
Note that the decomposition in the scalar sector of \eqref{IrredDec} is not unique. Alternatively, we could choose a basis in which the gauge invariant physical components become manifest.
We can eliminate the trace $h$ in favor of the conformal mode $\sigma$, which is defined as the gauge invariant combination,
\begin{align}
\sigma\coloneqq h+2\Delta b,\quad \delta_{\xi}\sigma=0\,.\label{ConformalMode}
\end{align}

\subsubsection{Fluctuation operator}
The irreducible decomposition \eqref{IrredDec} suggests the change of variables
\begin{align}
\tensor{h}{_\mu_\nu}\mapsto
\left(\tensor*{h}{^{\perp}_{\mu\nu}},\tensor*{v}{^{\perp}_{\mu}},h,b\right)^{\mathrm{T}}\label{IrrDecVec}\,,
\end{align}
where now each component is considered as an independent field.
The fluctuation operator on de Sitter space $\mathbf{F}_0$ then acquires the block matrix form
\begin{align}
\mathbf{F}_0=
\begin{bmatrix}
\mathbf{F}_{\mathrm{t}}&&&\\
&\mathbf{F}_{\mathrm{v}}&&\\
&&\mathbf{F}_{hh}&\mathbf{F}_{hb}\\
&&\mathbf{F}_{bh}&\mathbf{F}_{bb}\\
\end{bmatrix}
\,.\label{FlucOpBlock}
\end{align}
The individual components are given by
\begin{align}
\mathbf{F}_{\mathrm{t}}={}&-\frac{f_1}{2}\left(\Delta-\frac{1}{3}R_0+\frac{f}{f_1}\right)\,,\\
\mathbf{F}_{\mathrm{v}}={}&f_1\left(\Delta-\frac{3}{4}R_0+\frac{f}{f_1}\right)\left(\Delta-\frac{1}{4}R_0\right)\,,\\
\mathbf{F}_{hh}={}&\frac{9f_2}{16}\left[\Delta^2-\left(\frac{2}{3}R_0-\frac{2}{9}\frac{f_1}{f_2}\right)\Delta\right.\nonumber\\
&\phantom{\frac{9\,f_2}{16}}\;\,\left.+\frac{R_0^2}{9}-\frac{2}{9}\frac{f_1}{f_2}R_0+\frac{2}{9}\frac{f}{f_2}\right]\,,\\
\mathbf{F}_{bb}={}&\frac{9f_2}{4}\left[\Delta^2-\left(\frac{1}{3}R_0+\frac{2}{3}\frac{f_1}{f_2}\right)\Delta\right.\nonumber\\
&\phantom{\frac{9f_2}{4}}\;\,\left.+\frac{2}{3}\frac{f_1}{f_2}R_0-\frac{2}{3}\frac{f}{f_2}\right]\left[\Delta-\frac{1}{3}R_0\right]\Delta\,,\\
\mathbf{F}_{hb}={}&\mathbf{F}_{bh}=\frac{9f_2}{8}\left(\Delta-\frac{1}{3}R_0\right)^2\Delta\,.
\end{align}
The change of variables for differentially constrained fields leads to
additional functional determinants from the Jacobian $\mathbf{J}_h$, which is extracted from
\begin{align}
\langle h, h\rangle_2={}&\langle h^\perp, h^\perp\rangle_2+\left\langle v^{\perp},\left(\Delta-\frac{1}{4}R_0\right)v^{\perp}\right\rangle_1\nonumber\\
&-\frac{1}{8}\left\langle h,h\right\rangle_0+\frac{3}{2}\left\langle b,\Delta\left(\Delta-\frac{1}{3}R_0\right)b\right\rangle_0\,.
\end{align}
The Jacobian block operator $\mathbf{J}_h$, understood as acting on a vector \eqref{IrrDecVec}, reads
\begin{align}
\mathbf{J}_h=\begin{bmatrix}
\mathbf{1}&&&\\
&\Delta-\frac{1}{4}R_0&&\\
&&1&\\
&&&\left(\Delta-\frac{1}{3}R_0\right)\,\Delta\\
\end{bmatrix}\,.\label{JacobianOP}
\end{align}

\subsubsection{Ghost operator}
A similar decomposition is carried out for the ghost sector. The (anti)ghost fields decompose as
\begin{align}
\omega^{\mu}={}&\omega^{\mu}_{\perp}+\nabla ^{\mu}u\,,\quad \tensor{\nabla}{_\mu}\omega^{\mu}_{\perp}=0\,,\label{GhostDec}\\
\overline{\omega}_{\mu}={}&\overline{\omega}_{\mu}^{\perp}+\tensor{\nabla}{_\mu}\overline{u}\,,\quad \tensor{\nabla}{^\mu}\overline{\omega}_{\mu}^{\perp}=0\label{AntiGhost}\,.
\end{align}
In analogy to the previous subsection, we find for the block matrices of the ghost operator and ghost Jacobian
\begin{align}
\mathbf{Q}_0=\begin{bmatrix}
\Delta-\frac{1}{4}R_0&\\
&\Delta\left(\Delta-\frac{1}{2}R_0\right)\\
\end{bmatrix}\,,\;\;
\mathbf{J}_\omega=\begin{bmatrix}
\mathbf{1}&\\
&\Delta\\
\end{bmatrix}\,.\label{OPGhDec}
\end{align}

\subsubsection{Evaluation of traces}
The one-loop divergences for \eqref{FlucOpBlock} decompose into the sum of traces over transverse-traceless tensors, transverse vectors and scalars:
\begin{align}
\left. \operatorname{Tr}\ln \mathbf{F}_0 \right|^{\mathrm{div}}={}&\left. \operatorname{Tr}_{2,\perp}\ln \mathbf{F}_{\mathrm{t}} \right|^{\mathrm{div}}+
\left. \operatorname{Tr}_{1,\perp}\ln\mathbf{F}_{\mathrm{v}} \right|^{\mathrm{div}}\nonumber\\
&+\left.\operatorname{Tr}\ln
\begin{bmatrix}
\mathbf{F}_{hh}&\mathbf{F}_{hb}\\
\mathbf{F}_{bh}&\mathbf{F}_{bb}\\
\end{bmatrix}\right|^{\mathrm{div}}\,.\label{TrlnDivFIrred}
\end{align}
The transversal-traceless tensor trace and transversal vector traces are given by
\begin{align}
\left. \operatorname{Tr}_{2,\perp}\ln \mathbf{F}_{\mathrm{t}} \right|^{\mathrm{div}}={}&\left. \operatorname{Tr}_{2,\perp}\ln\left(\Delta-\frac{1}{3}R_0+\frac{f}{f_1}\right) \right|^{\mathrm{div}}\,,\label{TTTensorTrace}\\
\left. \operatorname{Tr}_{1,\perp}\ln\mathbf{F}_{\mathrm{v}} \right|^{\mathrm{div}}
={}&\left. \operatorname{Tr}_{1,\perp}\ln\left(\Delta-\frac{3}{4}R_0+\frac{f}{f_1}\right) \right|^{\mathrm{div}}\nonumber\\
&+\left. \operatorname{Tr}_{1,\perp}\ln\left(\Delta-\frac{1}{4}R_0\right) \right|^{\mathrm{div}}\,.\label{PrimeTrVector}
\end{align}
On a de Sitter background, all operators in the scalar trace commute with each other and we evaluate the trace of the scalar block operator as
\begin{align}
\operatorname{Tr}\ln
\begin{bmatrix}
\mathbf{F}_{hh}&\mathbf{F}_{hb}\\
\mathbf{F}_{bh}&\mathbf{F}_{bb}\\
\end{bmatrix}={}\operatorname{Tr}_0\ln\left(\mathbf{F}_{hh}\mathbf{F}_{bb}-\mathbf{F}_{hb}\mathbf{F}_{bh}\right)\,.\label{FsDef}
\end{align}
Here, ${\mathbf{F}_{\mathrm{s}}\coloneqq{} \mathbf{F}_{hh}\mathbf{F}_{bb}-\mathbf{F}_{hb}\mathbf{F}_{bh}}$ is a scalar operator of order ten.
The scalar trace in \eqref{FsDef} decomposes into the following sum of scalar traces:
\begin{align}
\left. \operatorname{Tr}_0\ln \mathbf{F}_{\mathrm{s}} \right|^{\mathrm{div}}=&\left. \operatorname{Tr}_0\ln\Delta \right|^{\mathrm{div}}+\left. \operatorname{Tr}_0\ln\left(\Delta-\frac{1}{3}R_0\right) \right|^{\mathrm{div}}\nonumber\\
&+\left. \operatorname{Tr}_0\ln\left(\Delta-R_0+\frac{f}{f_1}\right) \right|^{\mathrm{div}}\nonumber\\
&+\operatorname{Tr}_0\ln\left[\Delta^2-\frac{1}{3}\left(\frac{5}{2}R_0-\frac{f_1}{f_2}\right)\Delta\right.\nonumber\\
&\left. \left.+\frac{1}{3}\left(\frac{1}{2}R_0^2-\frac{f_1}{f_2}R_0+\frac{f}{f_2}\right)\right] \right|^{\mathrm{div}}\,.\label{PrimeTraceScalar}
\end{align}
Note that the last scalar trace is identical to the first trace in \eqref{TrlogDivScalConstCurvOff}.
Similarly, the divergent contribution from the Jacobian \eqref{JacobianOP} is given by
\begin{align}
\left. \operatorname{Tr}\ln\mathbf{J}_h \right|^{\mathrm{div}}={}&\left. \operatorname{Tr}_{1,\perp}^{\prime}\ln\left(\Delta-\frac{1}{4}R_0\right) \right|^{\mathrm{div}}
+\left. \operatorname{Tr}^{\prime}_0\ln\Delta \right|^{\mathrm{div}}\nonumber\\
&+\left. \operatorname{Tr}_0^{\prime}\ln\left(\Delta-\frac{1}{3}R_0\right) \right|^{\mathrm{div}}\,.\label{TrLnJacobian}
\end{align}
A prime on a trace indicates the subtraction of the modes associated with the lowest eigenvalue. Apart from these modes, there is a cancellation of contributions from the fluctuation operator \eqref{PrimeTrVector} and \eqref{PrimeTraceScalar}  with contributions from \eqref{TrLnJacobian}.
The ghost trace decomposes as
\begin{align}
\left. \operatorname{Tr}\ln\mathbf{Q}_0 \right|^{\mathrm{div}}={}&\left. \operatorname{Tr}\ln_{1,\perp}\left(\Delta-\frac{1}{4}R_0\right) \right|^{\mathrm{div}}+\left. \operatorname{Tr}_0\ln\Delta \right|^{\mathrm{div}}\nonumber\\
&+\left. \operatorname{Tr}_0\ln\,\left(\Delta-\frac{1}{2}R_0\right) \right|^{\mathrm{div}}\,.\label{TRLogDivQBlock}
\end{align}
Again, the contribution of the Jacobian cancels one of the traces in \eqref{TRLogDivQBlock} up to zero modes:
\begin{align}
\left. \operatorname{Tr}\ln \mathbf{J}_\omega \right|^{\mathrm{div}}=\left. \operatorname{Tr}_0^{\prime}\ln\Delta \right|^{\mathrm{div}}\,.\label{TrLnDivJacGhost}
\end{align}
The evaluation of the nontrivial traces in \eqref{TTTensorTrace}, \eqref{PrimeTrVector} and \eqref{PrimeTraceScalar} can be found in Appendix \ref{TracesIrredDec}.
Altogether, the divergent part of the one-loop effective action on a de Sitter background, obtained in terms of the irreducible decomposition, reads  
\begin{align}
\left. \varGamma_1 \right|^{\mathrm{div}}_{R_0}=&{}\frac{1}{2}\left. \operatorname{Tr}\ln\mathbf{F}_0 \right|^{\mathrm{div}}-\frac{1}{2}\left. \operatorname{Tr}\ln \mathbf{J}_h \right|^{\mathrm{div}}\nonumber\\
&-\left. \operatorname{Tr}\ln\mathbf{Q}_0 \right|^{\mathrm{div}}+\left. \operatorname{Tr}\ln \mathbf{J}_\omega \right|^{\mathrm{div}}\nonumber\\
=&{}\frac{1}{\varepsilon}\left[-\frac{313}{20}+\frac{1}{2}N_{\mathrm{tot}}-54\left(\frac{f}{R_0\,f_1}\right)^2+54\frac{f}{R_0\,f_1}\right.\nonumber\\
&\quad\,\left.-\frac{2}{3}\left(\frac{f_1}{R_0\,f_2}\right)^2+4\frac{f}{R_0^2f_2}\right]\,.\label{OneLoppDivIrred}
\end{align}
The total number of zero modes and negative modes is given by
\begin{align}
N_{\mathrm{tot}}=N(\mathbf{J}_h)-2 N(\mathbf{J}_\omega)=16-2=14\,,\label{DiscreteModes}
\end{align}
where the traces in \eqref{TrLnJacobian} contribute ten zero modes, five negative modes and one zero mode to $N(\mathbf{J}_h)$, respectively, while the trace in \eqref{TrLnDivJacGhost} contributes one zero mode to $N(\mathbf{J}_\omega)$; see e.g.\ Table 8 in \cite{Codello:2008vh}.\footnote{These modes are related to the symmetries of the de Sitter background; see e.g.\ \cite{Fradkin:1983mq, Allen:1986ta, Polchinski:1988ua, Taylor:1989ua, Vassilevich:1992rk, Mottola:1995sj, Volkov:2000ih} for more details.}
Inserting \eqref{DiscreteModes} into \eqref{OneLoppDivIrred}, the result coincides with \eqref{Gam1LoopdeSitter}.


\section{Checks and comparison with previous results}\label{Sec:CheckAndComp}

\subsection{Comparison with one-loop calculation for Einstein gravity with a cosmological constant}\label{SubSec:EinsteinGravCosConst}   
General relativity with a cosmological constant corresponds to the special case of \eqref{fRAct} with 
\begin{align}
f(R)=-\frac{M_\mathrm{P} ^2}{2}\left(R-2\Lambda\right)\,,
\end{align}
where $\Lambda$ is the cosmological constant and $M_{\mathrm{P}}$ the Planck mass. In particular, we have 
\begin{align}
f_2=0\,,\quad\tensor{\Upsilon}{_\mu}=0\,.\label{EinsteinCond}
\end{align}
As noted before, in this case the scalar contribution is absent from the divergent part of the effective action \eqref{TrLogDivf2Zero}, which allows us to test the tensor and vector contributions by comparing them to previous calculations performed in \cite{tHooft:1974toh, Christensen:1979iy, Barvinsky:1985an}.
Using \eqref{TrLogDivTensor} and \eqref{TrlogDivVec}, the result for the one-loop divergences reads
\begin{align}
\left. \varGamma_1 \right|^{\mathrm{div}}_{\mathrm{EH}}={}&\frac{1}{2}\left. \operatorname{Tr}_{2}\ln\mathbf{H} \right|^{\mathrm{div}}_{\mathrm{EH}}-\left. \operatorname{Tr}_{1}\ln\mathbf{Q} \right|^{\mathrm{div}}_{\mathrm{EH}}\nonumber\\
=&\frac{1}{16\pi^2\varepsilon}\int{\rm d}^4x\, g^{\nicefrac{1}{2}}\left(-\frac{53}{90}\mathcal{G}-\frac{7}{20}\tensor*{R}{_\mu_\nu}\tensor*{R}{^\mu^\nu}\right.\nonumber\\
&\phantom{\frac{1}{16\pi^2\varepsilon}\int}\left.-\frac{1}{120}R^2+\frac{13}{6}\Lambda R-\frac{5}{2}\Lambda^2\right)\,. \label{Gam1LoopEinsteinCC}
\end{align}
For $\Lambda= 0$, we recover the well-known result for Einstein gravity without a cosmological constant \cite{tHooft:1974toh, Barvinsky:1985an}.

For $\Lambda\neq0$, we compare with the calculation in \cite{Christensen:1979iy}, which is performed on an Einstein space. In view of \eqref{EOM} and \eqref{EinsteinCond} this is equivalent to the equation of motion
\begin{align}
\tensor*{R}{_\mu_\nu}=\Lambda\,\tensor{g}{_\mu_\nu}\,.
\end{align}
Therefore, on shell \eqref{Gam1LoopEinsteinCC} reduces to 
\begin{align}
\left. \varGamma_1 \right|^{\mathrm{div}}_{\mathrm{EH},\mathscr{E}=0}={}\frac{1}{\varepsilon}\left[-\frac{53}{45}\chi({\cal M})+\frac{87}{20}\frac{\Lambda^2 V({\cal M})}{12\pi^2}\right]\,,\label{Gam1LoopEinsteinCCOnShell}
\end{align}
where $V({\cal M})$ is the volume of  ${\cal M}$ and $\chi({\cal M})$ the Euler characteristic of ${\cal M}$,
\begin{align}
V({\cal M})\coloneqq{}&\int_{\cal M}\mathop{}\!\mathrm{d}^{4}{x}\, g^{\nicefrac{1}{2}}\,,\\
\chi({\cal M})\coloneqq{}&\frac{1}{32\pi^2}\int_{\cal M}\mathop{}\!\mathrm{d}^{4}{x}\, g^{\nicefrac{1}{2}}\,\mathcal{G}\,.
\end{align}
The result \eqref{Gam1LoopEinsteinCCOnShell} is in perfect agreement with \cite{Christensen:1979iy}.\footnote{Note that they present the poles in dimension as $d-4=-2\varepsilon$.}

\subsection{Comparison with zeta function calculation for $f(R)$ gravity on de Sitter space}\label{SubSubSec:CompPrevCalcdeSitter}
Since the spectrum of the Laplacian on a sphere is known explicitly, the one-loop effective action can be evaluated by the zeta function technique.
In combination with the irreducible decomposition \eqref{IrredDec}, this zeta function technique is used in \cite{Cognola:2005de} to calculate the one-loop effective action for $f(R)$ gravity on a de Sitter background. 
We extract the one-loop divergences from the result presented in \cite{Cognola:2005de}, Eq. (3.33), by focusing on the contributions proportional to $\ln (\ell^2R_0/12)$.
Here, $\ell$ is a reference scale with dimension of length.
Since the authors have chosen a different gauge, we compare the gauge-independent on shell result.
Inserting the logarithmic contributions from Eqs. (B.29), (B.35) and (B.38) into Eq. (3.33) and using the on shell relation \eqref{OnShellCondConstCurv}, we find
\begin{align}
\left. \varGamma_1 \right|^{\mathrm{div}}_{\mathscr{E}_0=0}=\frac{1}{\varepsilon}&\left[-\left(\frac{143}{20}+\frac{1}{2}N_0^{\mathrm{tot}}\right)\right.\nonumber\\
&\left.\;\,+ 4\,\frac{f}{R_0^2f_2}-\frac{8}{3}\,\left(\frac{f}{R_0^2f_2}\right)^2\right]\,.\label{NojiriOnShellZeroModes}
\end{align}
According to \cite{Cognola:2005de}, $N_0^{\mathrm{tot}}$ is the total number of zero modes.\footnote{Ten zero modes from the vector sector are already taken into account in the result \eqref{NojiriOnShellZeroModes}.}
The $f$-dependent structures coincide with our on shell result \eqref{Gam1LoopOnShelldeSitter}. However, the $f$-independent structure seems to be incompatible with \eqref{Gam1LoopOnShelldeSitter}, as it would require ${N_0^{\mathrm{tot}}=-24}$.

\subsection{Comparison with the $f(R)$ truncation in the functional renormalization group}
We recover the one-loop result on a de Sitter background from the functional renormalization group flow of $f(R)$ gravity similar to the procedure described e.g. in \cite{Codello:2015oqa,Codello:2017hhh,Merzlikin:2017zan}. 
The Wetterich equation describes the functional renormalization group flow of the effective averaged action $\varGamma_k$ with respect to the momentum scale $k$ \cite{Wetterich:1992yh, Reuter:1993kw},
\begin{align}
\partial_t\varGamma_k=\frac{1}{2}\operatorname{Tr}\left(\frac{\delta^2\varGamma_k}{\delta\phi\,\delta\phi}+R_k\right)^{-1}\partial_tR_k\,.\label{Wettericheq}
\end{align}
Here $t=\ln(k/\mu)$ is the logarithmic scale with the arbitrary reference scale $\mu$ and $R_k$ is a scale-dependent regulator function.
In the one-loop approximation the Wetterich equation reduces to \cite{Codello:2008vh},
\begin{align}
\partial_t\left(S+\varGamma_k^{1}\right)=\frac{1}{2}\operatorname{Tr}\left(\frac{\delta^2S}{\delta\phi\,\delta\phi}+R_k\right)^{-1}\partial_tR_k\,.\label{WetterichOneLoop}
\end{align}
This corresponds to a replacement of the full effective action $\varGamma_k$ by its one-loop approximation $S+\varGamma^{1}_k$ on the left-hand side of \eqref{Wettericheq} and a replacement of the effective action $\varGamma_k$ by the ``bare'' action on the right-hand side of \eqref{Wettericheq}. 
In \cite{Codello:2008vh}, the functional trace in \eqref{Wettericheq} is calculated for the $f(R)$ truncation.
Equations (113)--(114) in \cite{Codello:2008vh} are expressed in terms of the dimensionless variables
\begin{align}
\widetilde{R}=R\,k^{-2}\,,\quad \widetilde{f}_n=f_n\,k^{2(n-2)}\,.
\end{align} 
We extract the one-loop result from Eq.\ (113) in \cite{Codello:2008vh} by neglecting the explicit scale dependence of $\widetilde{f}$, that is by setting
\begin{align}
\partial_t\widetilde{f}_n=0\,.
\end{align}
Restoring the original dimensionful quantities $R$ and $f_n$ by introducing explicit factors of the momentum scale $k$, we integrate the flow from the UV scale $k=\Lambda$ down to the reference scale $k=\mu$ and obtain the one-loop approximation of the effective action:
\begin{align}
\varGamma^{1}=-\int_{\mu}^\Lambda\mathop{}\!\mathrm{d} k\,\partial_k\varGamma_{k}^1\,.
\end{align}
The resulting expression for the effective action contains contributions that diverge as $\Lambda\to\infty$,
\begin{align}
\varGamma^{1}=\varGamma^{1}_{\mathrm{quart}}+\varGamma^{1}_{\mathrm{quad}}+\varGamma^{1}_{\mathrm{log}}+\text{UV-finite terms}\,.
\end{align}
Here,  $\varGamma^{1}_{\mathrm{quart}},\varGamma^{1}_{\mathrm{quad}},\varGamma^{1}_{\mathrm{log}}$ are quartic, quadratic and logarithmic divergent contributions respectively. As dimensional regularization annihilates all power law divergences, only the logarithmic divergent part is relevant for the comparison to our result. 
It is isolated by expanding the integrand $\partial_k\varGamma_{k}^1$ around $k=\infty$, and extracting the terms proportional to $k^{-1}$. In this way we find
\begin{align}
\varGamma^{1}_{\mathrm{log}}={}&\ln\left(\frac{\Lambda^2}{\mu^2}\right)\left[\frac{147}{20}-30\left(\frac{f}{R_0\,f_1}\right)^2+10\frac{f}{R_0f_1}\right.\nonumber\\
&\left.-\frac{2}{3}\left(\frac{f_1}{R_0f_2}\right)^2-\frac{8}{3}\frac{f}{R_0^2f_2}+\frac{2}{3}\frac{f_1}{R_0f_2}\right]\,.
\end{align}
On shell, this reduces to 
\begin{align}
\varGamma^{1}_{\mathrm{log},\mathscr{E}_0=0}=\ln\left(\frac{\Lambda^2}{\mu^2}\right)\left[\frac{97}{20}+4\frac{f}{R_0^2f_2}-\frac{8}{3}\left(\frac{f}{R_0^2f_2}\right)^2\right]\,.
\end{align}
Identifying  $\ln \left(\Lambda^2/\mu^2\right)=1/\varepsilon$, this expression agrees with our on shell result on de Sitter space \eqref{Gam1LoopOnShelldeSitter}.\footnote{The on shell agreement is obtained only for the ``$\alpha$-gauge'', corresponding to the second expression for $\Sigma$ in Eq.\ (114) in \cite{Codello:2008vh} [the factor of $384\pi^2$ in Eq.\ (113) in \cite{Codello:2008vh} is presumably a typo and set to $1$ in our comparison].}


\section{Conclusion}\label{Sec:Conclusion}
Our main result is the calculation of the divergent part of the one-loop effective action for $f(R)$ gravity on an arbitrary background. 
This generalizes previous calculations of quantum corrections in $f(R)$ gravity, which have been restricted to constant curvature backgrounds. Allowing for arbitrary backgrounds increases the complexity of the one-loop calculation considerably, but permits us to access the individual coefficients of the quadratic curvature invariants and the structures involving derivatives of the Ricci scalar.

Our result is relevant for cosmological $f(R)$ theories, as it allows to investigate the influence of quantum corrections on the dynamics of a time-dependent Friedmann-Robertson-Walker background.
Other interesting applications include the study of black hole solutions in $f(R)$ gravity. 
Note, that on shell the derivative structures in \eqref{Gamma1LoopOffShellFinalGamPara} also contribute to the renormalization of the $f(R)$ structure.
On constant curvature backgrounds these contributions are absent as they are proportional to algebraic on shell condition $\mathscr{E}_0$ and vanish. 

We have represented our main result in terms of three different bases in the space of invariants in order to perform the on shell reduction and de Sitter reduction.
As an internal consistency check of our method, we have repeated the calculation directly in de Sitter space---once with the same method as for the calculation on an arbitrary background, and once by a decomposition of the fluctuation field into its irreducible components.
All results on de Sitter space are consistent with the de Sitter reduction of the result for arbitrary backgrounds.
In addition, we have compared our result with previous calculations in $f(R)$ gravity. 
We find perfect agreement with the one-loop calculations for Einstein gravity \cite{tHooft:1974toh,Barvinsky:1985an} and Einstein gravity with a cosmological constant \cite{Christensen:1979iy}.
We also coincide with the exact renormalization group analysis, obtained for the $f(R)$ truncation on a de Sitter background \cite{Codello:2008vh}. 
Apart from the coefficient for the $f$-independent structure, we also agree with the one-loop result for $f(R)$ gravity in de Sitter space of \cite{Cognola:2005de}.   

Besides the main result, our calculation contains an interesting technical aspect. 
Standard perturbative heat-kernel methods for UV divergences are based on an expansion around the principal part of the fluctuation operator.
These methods fail if the principal symbol is degenerate, which is the case for $f(R)$ gravity where the degeneracy is a result of the dyadic structure of the principal part.
The technique developed in this article turns this drawback into an advantage. In fact, as discussed in detail in Sec. \ref{SubSec:DegPrincipalPart}, the dyadic structure is the essential element of our technique and allows us to complete the calculation in the first place.
Moreover, it naturally organizes the calculation in a transparent way, by subdividing it into the evaluation of three functional traces \eqref{Gam1Full} and thereby isolates the contribution of the additional higher-derivative scalar degree of freedom.

Finally, since our result captures all one-loop structures, it provides the basis for an investigation of the equivalence between $f(R)$ gravity and scalar-tensor theories at the one-loop quantum level, which we address in a separate publication \cite{WeFrameComparison}.

\begin{acknowledgments}
M. S. R. acknowledges financial support from the Deutschlandstipendium. 
\end{acknowledgments}


\appendix
\allowdisplaybreaks[1]
\section{Formalism and Notations}\label{Sec:NotationFormalism}

\subsection{Bundle structure}\label{SubSec:BundleStructure}
The formalism involves the general structure of a vector bundle ${\cal V}$ over a $d=4$ dimensional Riemannian manifold $({\cal M},g)$.
Fields $\phi$ are elements of the space of smooth sections ${\cal F}=C^{\infty}({\cal V})$ of ${\cal V}$.
In local coordinates we identify a field $\phi$ by its components $\phi^{A}(x)$, where $A,B,\dots$ are the bundle indices,
\begin{align}
\phi\mapsto\phi^{A}(x)\,.
\end{align}
In addition, we assume that ${\cal V}$ is endowed with a metric compatible, torsion-free affine connection $\nabla$.
Throughout the paper, we denote matrix valued operators ${\mathbf{L}\colon {\cal F}_{1}\to{\cal F}_2}$ in boldface:
\begin{align}
\mathbf{L}\mapsto\tensor{L}{^A_B}.
\end{align}
On the space of fields ${\cal F}$, we define an inner product 
\begin{align}
&\langle\phi_1,\,\phi_2\rangle
=\int\mathop{}\!\mathrm{d}^{4}x\, g^{\nicefrac{1}{2}}\, \phi^A_1(x)\,\tensor{\gamma}{_{AB}}(x)\,\phi^B_2(x)\,,\label{IPGen}
\end{align}
where the bundle metric $\tensor{\gamma}{_A_B}$ satisfies  
\begin{align}
\det \tensor{\gamma}{_{AB}}\neq0\,,\quad \tensor{\gamma}{_{AC}}\tensor{\gamma}{^{CB}}=\tensor*{\delta}{^{B}_{A}}.
\end{align} 
For an operator ${\mathbf{L}\colon{\cal F}_{1}\to{\cal F}_2}$, the inner products on ${\cal F}_1$ and ${\cal F}_2$ allow us to define the (formal) adjoint ${\mathbf{L}^{\dagger}\colon{\cal F}_2\to{\cal F}_1}$, 
\begin{align}
\langle\phi_2,\mathbf{L}\,\phi_1\rangle_{2}=\langle\mathbf{L}^{\dagger}\,\phi_2,\,\phi_1\rangle_{1}\,,\quad\phi_1\in{\cal F}_{1},\; \phi_{2}\in{\cal F}_2\,.\label{AdjointInnerProd}
\end{align}
Here, $\mathbf{1}\mapsto \tensor*{\delta}{^{A}_{B}}$ is the identity operator and $\tensor{\mathscr{R}}{_\mu_\nu}\mapsto\tensor{\mathscr{R}}{_\mu_\nu^A_B}$ is the bundle curvature, defined by the commutator
\begin{align}
\left[\tensor{\nabla}{_\mu},\tensor{\nabla}{_\nu}\right]\tensor{\phi}{^A}=\tensor{\mathscr{R}}{_\mu_\nu^A_B}\tensor{\phi}{^B}\,.
\end{align}
The reduction of the general formalism to the case of symmetric rank-two tensor fields $\tensor{h}{_\mu_\nu}$, vector fields $\xi^{\mu}$ and scalar fields $\varphi$ is summarized in Table \ref{Table1}. 
\begin{widetext}
\begin{center}
\begin{table}[H]
\begin{ruledtabular}
\begin{tabular}{cccccccl}
    ${\cal F}$& $\phi^{A}$& $\tensor{\gamma}{_{AB}}$& $\tensor{\gamma}{^{AB}}$& $\delta^{A}_{B}$& $\operatorname{tr}\delta^{A}_{B}$& $[\mathscr{R}_{\mu\nu}]^{A}_{\;\;B}$ &\multicolumn{1}{c}{$\langle\phi_1,\,\phi_2\rangle$}\rule{0pt}{2.6ex}\rule[-1.2ex]{0pt}{0pt}\\
    \hline
    ${\cal F}^{2}$& $\tensor{h}{_\mu_\nu}$& $\tensor{\gamma}{^{\mu\nu,\rho\sigma}}$& $\tensor{\gamma}{_{\mu\nu,\rho\sigma}}$& $\delta^{\mu\nu}_{\rho\sigma}$& $10$& $2\tensor{R}{_\mu_\nu_{(\rho}^{(\alpha}}\tensor*{\delta}{^{\beta)}_{\sigma)}}$& $\langle h^{1},h^{2}\rangle_2=\int\mathop{}\!\mathrm{d}^{4}x\, g^{\nicefrac{1}{2}} h^{1}_{\mu\nu}\tensor{\gamma}{^{\mu\nu,\rho\sigma}}h^{2}_{\rho\sigma}$\\
    ${\cal F}^{1}$& $\xi^{\mu}$& $\tensor{g}{_\mu_\nu}$& $\tensor{g}{^\mu^\nu}$& $\delta^{\mu}_{\nu}$& $4$& $\tensor{R}{_{\mu\nu}^\alpha_\beta}$ &$\langle \xi_1,\xi_2\rangle_1=\int\mathop{}\!\mathrm{d}^{4}x\, g^{\nicefrac{1}{2}} \,\xi_{1}^{\mu}\tensor{g}{_\mu_\nu}\xi_{2}^{\nu}$\\
    ${\cal F}^{0}$& $\varphi$& $1$& $1$& $1$& $1$& $0$& $\langle\varphi_{1},\varphi_{2}\rangle_0=\int\mathop{}\!\mathrm{d}^{4}x\, g^{\nicefrac{1}{2}}\varphi_1\,\varphi_2$\\
\end{tabular}
\end{ruledtabular}
\caption{Reduction of the formalism for generalized fields $\phi^{A}$ to rank-two symmetric tensor fields $\tensor{h}{_\mu_\nu}$, vector fields $\xi^{\mu}$ and scalar fields $\varphi$.}
\label{Table1}
\end{table}
\end{center}
\end{widetext}
\noindent The explicit expression for the ultralocal, dedensitized DeWitt metric $\tensor{\gamma}{^{\mu\nu,\rho\sigma}}$ and its inverse $ \tensor{\gamma}{_{\mu\nu,\rho\sigma}}$ are 
\begin{align}
\tensor{\gamma}{^{\mu\nu,\rho\sigma}}={}&\frac{1}{4}\left(\tensor{g}{^\mu^\rho}\tensor{g}{^\nu^\sigma}+\tensor{g}{^\mu^\sigma}\tensor{g}{^\nu^\rho}-\tensor{g}{^\mu^\nu}\tensor{g}{^\rho^\sigma}\right),\\
\tensor{\gamma}{_{\mu\nu,\rho\sigma}}={}& \tensor{g}{_\mu_\rho}\tensor{g}{_\nu_\sigma}+\tensor{g}{_\mu_\sigma}\tensor{g}{_\nu_\rho}-\tensor{g}{_\mu_\nu}\tensor{g}{_\rho_\sigma},\label{InvBundleMetric}\\
\tensor{\gamma}{^{\mu\nu,\alpha\beta}} \tensor{\gamma}{_{\alpha\beta,\rho\sigma}}={}&\delta^{\mu\nu}_{\rho\sigma}=\frac{1}{2}\left(\tensor*{\delta}{^{\mu}_{\rho}}\tensor*{\delta}{^{\nu}_{\sigma}}+\tensor*{\delta}{^{\mu}_{\sigma}}\tensor*{\delta}{^{\nu}_{\rho}}\right)\,.
\end{align}

\subsection{One-loop divergences}\label{SubSec:OneLoopDiv}
The fundamental fluctuation operator $\mathbf{F}\mapsto \tensor{F}{^{A}_{B}}(\nabla)$ is defined as the (formally) self-adjoint operator 
\begin{align}
\tensor{F}{^A_B}(\nabla_{x})\,\delta(x,x^\prime)\coloneqq g^{\nicefrac{-1}{2}}\,\tensor{\gamma}{^A^C}\frac{\delta^2 S_{\mathrm{tot}}}{\delta\phi^C(x)\,\delta\phi^{B}(x^{\prime})}\,.\label{FlucOp}
\end{align}
Note that the delta function $\delta(x,x')$ has density weight zero in the first argument $x$ and unit weight at $x'$.  
The one-loop effective action is expressed as the sum of three functional traces for the gauge-fixed fluctuation operator $\mathbf{F}$, the ghost operator $\mathbf{Q}$ and the Nielsen-Kallosh operator $\mathbf{O}$, 
\begin{align}
\varGamma_1=\frac{1}{2}\operatorname{Tr}\ln\mathbf{F}-\operatorname{Tr}\ln \mathbf{Q}+\frac{1}{2}\operatorname{Tr}\ln \mathbf{O}\,.\label{OneLoopGeneral}
\end{align}
The ghost contribution enters twice with opposite sign, due to the Grassmannian nature of the ghost field and the antighost field. 
The Nielsen-Kallosh operator only gives a nontrivial contribution to the one-loop divergences in case it involves derivatives. 
The functional trace of a two point tensor $T^{A}_{B}(x,y)$ involves integration over its coincidence limit and the internal bundle trace $\operatorname{tr}$,
\begin{align}
\operatorname{Tr}  \tensor{T}{^{A}_{B}}(x_A,x_B)={}&\int\mathop{}\!\mathrm{d}^{4}x\,\operatorname{tr} \tensor{T}{^{A}_{B}}(x,x)\nonumber\\
={}&\int\mathop{}\!\mathrm{d}^{4}x\, \tensor{T}{^{A}_{A}}(x)\,.
\end{align}
For a minimal second-order operator with potential $\mathbf{P}$,
\begin{align}
\mathbf{L}=\Delta+\mathbf{P},\quad \Delta\coloneqq-\tensor{\nabla}{^\mu}\tensor{\nabla}{_\mu}\,,
\label{MinSecOp}
\end{align}
the divergent part of the functional trace $\operatorname{Tr}\ln \mathbf{L}$ can be calculated by the heat-kernel-based Schwinger-DeWitt technique \cite{DeWitt:1965, Barvinsky:1985an}. 
In dimensional regularization divergences are isolated as poles in dimension $\varepsilon=2-d/2$ for $d\to4$.
The one-loop divergences can be given in closed form \cite{DeWitt:1965},
\begin{align}
\left. \operatorname{Tr}\ln\mathbf{L} \right|^{\mathrm{div}}=-\frac{1}{16\pi^2\varepsilon}\int{\rm d}^4x\, g^{\nicefrac{1}{2}}\operatorname{tr}\mathbf{a}_{2}\,.
\label{TRLNDiv}
\end{align} 
Here, $\mathbf{a}_2$ is the coincidence limit $x'\to x$ of the second Schwinger-DeWitt coefficient
\begin{align}
\mathbf{a}_{2}={}&\frac{1}{180}\left(\tensor{R}{_{\mu\nu\rho\sigma}}\tensor{R}{^{\mu\nu\rho\sigma}}-\tensor*{R}{_{\mu\nu}}\tensor*{R}{^{\mu\nu}}\right)\mathbf{1}\nonumber\\
&+\frac{1}{12}\tensor{\mathscr{R}}{_\mu_\nu}\tensor{\mathscr{R}}{^\mu^\nu}+\frac{1}{2}\left(\mathbf{P}-\frac{R}{6}\,\mathbf{1}\right)^2+\text{t. d.}\,,
\end{align}
where t.d. denotes total derivatives.

\subsection{Universal functional traces}\label{AppUFT}
Any differential operator $\mathbf{L}$ can be represented as a sum of terms ordered according to the number of derivatives:
\begin{align}
\mathbf{L}(\nabla)=\mathbf{D}(\nabla)+\mathbf{\Pi}(\nabla)\,.
\end{align}
Here, the principal part $\mathbf{D}$ encompasses the highest derivative part, while all lower order derivative terms are collected in the differential operator $\mathbf{\Pi}$. The components of the principal part read
\begin{align}
\mathbf{D}(\nabla)\mapsto[\tensor{D}{^{A}_{B}}]^{\mu_1\dots\mu_{2k}}\nabla_{\mu_1}\dots\nabla_{\mu_{2k}}\,.
\end{align}
We call an operator $\mathbf{L}$ minimal if its principal part is given by $[\tensor{D}{^{A}_{B}}]^{\mu_1\dots\mu_{2k}}\nabla_{\mu_1}\dots\nabla_{\mu_{2k}}=\tensor{D}{^{A}_{B}}\Delta^k$; otherwise we call it nonminimal.
For nonminimal or higher order operators, the original Schwinger-DeWitt algorithm for minimal second-order operators \eqref{MinSecOp} has to be modified.
In \cite{Barvinsky:1985an} the authors propose a generalization of the Schwinger-DeWitt algorithm, which is based on a perturbative expansion in $\mathbf{\Pi}$. Essential for this perturbative treatment is the notion of background dimension $\mathfrak{M}$, which is understood as the mass dimension of the background tensorial coefficients of the differential operator. We write $\mathbf{L}={\cal O}(\mathfrak{M}^k)$ for an operator $\mathbf{L}$, which has at least background dimension $\mathfrak{M}^k$. 
The generalized Schwinger-DeWitt algorithm allows to reduce the calculation of divergences for nonminimal and higher order operators to the evaluation of a few tabulated universal functional traces (UFT) \cite{Barvinsky:1985an} for the second-order minimal operator $\Delta+\mathbf{P}$,
\begin{align}
\tensor*{{[\tensor{\mathscr{U}}{^{A}_{B}}(P)]}}{^{(p,n)}_{\mu_1\dots\mu_p}}=\nabla_{\mu_1}\dots\nabla_{\mu_p}\frac{\tensor*{\delta}{^A_B}}{(\Delta+P)^n}\Big|^{\mathrm{div}}_{x^\prime=x}\,.
\end{align}
We denote the inverse of an operator $\mathbf{L}$ by $\mathbf{1}/L$, such that its bundle structure is indicated by the corresponding identity matrix $\mathbf{1}\mapsto\tensor*{\delta}{^A_B}$.
Different traces are characterized by the pair $(p,n)$ and can be classified according to their degree of divergence:
\begin{align}
\chi_\mathrm{div}=p-2n+d\,.
\end{align}
In $d=4$, divergent contributions arise for $0\leq\chi_{\mathrm{div}}\leq4$.
For $\mathbf{P}=0$, the UFT were introduced and tabulated in \cite{Jack:1983sk, Barvinsky:1985an}. These traces were extended for $\mathbf{P}\neq0$ in \cite{Barvinsky:1988ds}. We list all UFT with $\mathbf{P}=0$ appearing in our calculation:
\begin{align}
\mathscr{U}^{(2,3)}_{\mu\nu}={}&\frac{ g^{\nicefrac{1}{2}}}{16\pi^2\varepsilon}\,\left(-\frac{1}{4}\tensor{g}{_\mu_\nu}\right)\,,\\
\mathscr{U}^{(4,4)}_{\mu\nu\rho\sigma}={}&\frac{ g^{\nicefrac{1}{2}}}{16\pi^2\varepsilon}\,\frac{1}{24}\left(\tensor{g}{_\mu_\rho}\tensor{g}{_\nu_\sigma}+\tensor{g}{_\mu_\sigma}\tensor{g}{_\nu_\rho}+\tensor{g}{_\mu_\nu}\tensor{g}{_\rho_\sigma}\right)\,,\label{UFT1}\\
\mathscr{U}^{(3,3)}_{\mu\nu\rho}={}&0\,,\\
\mathscr{U}^{(0,1)}={}&\frac{ g^{\nicefrac{1}{2}}}{16\pi^2\varepsilon}\,\frac{1}{6}R\,,\\
\mathscr{U}^{(2,2)}_{\mu\nu}={}&\frac{ g^{\nicefrac{1}{2}}}{16\pi^2\varepsilon}\,\frac{1}{6}\left(\tensor*{R}{_\mu_\nu}-\frac{1}{2}\tensor{g}{_\mu_\nu}R\right)\,.
\label{UFTs}
\end{align}
Note that since all the scalar UFT listed above have ${\mathbf{P}=0}$, we have suppressed the argument $P$.

\onecolumngrid
\section{Explicit coefficients of the operators $\mathbf{B}$ and $\mathbf{X}$}\label{AppCoeffBX}

\subsection{Coefficients of the operator $\mathbf{B}$}\label{AppCoeffB}

Below, we list the tensorial background coefficients of the operator $\mathbf{B}$ as defined in \eqref{BCoeff}. Without loss of generality, we have defined the $B_i$ as totally symmetric tensors.\footnote{We use the \texttt{Mathematica} tensor algebra bundle \texttt{xAct} \cite{xAct,Nutma:2013zea,Brizuela:2008ra} to check the explicit coefficients \eqref{AppCoefB2}--\eqref{AppCoefX4}.}
Note that all coefficients $B_i$ are identically zero on a space of constant curvature such that $\mathbf{B}_0\equiv0$. Therefore, the operator identity \eqref{AGId} essentially simplifies in de Sitter space. In $d=4$, the coefficients read
\begin{align}
\tensor*{B}{^\mu^\nu^\rho^\sigma_2}={}&\left(\tensor{g}{^\alpha^\beta}\tensor{g}{^\rho^\sigma}
-\tensor{g}{^\alpha^{(\rho}}\tensor{g}{^{\sigma)}^\beta}\right)\left[\tensor{P}{_{\alpha\beta}^{\mu\nu}}-\left(\frac{f}{f_1}-R\right)\tensor*{\delta}{^{\mu\nu}_{\alpha\beta}}
+2\tensor{R}{^{\mu}_{(\alpha}^{\nu}_{\beta)}}\right]\nonumber\\
&+\tensor{g}{^\mu^{(\rho}}\left(\tensor*{R}{^{\sigma)}^\nu}+\tensor{\Upsilon}{^{\sigma)}^{;\nu}}\right)
+\tensor{g}{^\nu^{(\sigma}}\left(\tensor*{R}{^{\rho)}^\mu}+\tensor{\Upsilon}{^{;\rho)}^\mu}\right)-2\tensor{g}{^\mu^\nu}\left(\tensor*{R}{^\rho^\sigma}+\tensor{\Upsilon}{^{\rho}^{;\sigma}}\right)\nonumber\nonumber\\
={}&\tensor{g}{^\mu^{(\rho}}\tensor{g}{^{\sigma)}^\nu}\left(\frac{7}{4}\tensor{\Upsilon}{^\alpha}\tensor{\Upsilon}{_\alpha}-R+\frac{3}{2}\tensor{\Upsilon}{_\alpha^{;\alpha}}\right)+\tensor{g}{^\mu^\nu}\tensor{g}{^\rho^\sigma}\left(-\frac{5}{4}\tensor{\Upsilon}{^\alpha}\tensor{\Upsilon}{_\alpha}+\frac{1}{2}R-\tensor{\Upsilon}{_\alpha^{;\alpha}}\right)\nonumber\\
&-\tensor{g}{^\rho^\sigma}\left(\tensor*{R}{^\mu^\nu}-\tensor{\Upsilon}{^\mu}\tensor{\Upsilon}{^\nu}-2\tensor{\Upsilon}{^{\mu;\nu}}\right)\nonumber\\
&+2\tensor{g}{^\mu^{(\rho}}\left(\tensor{R}{^{\sigma)}^{\nu}}-\tensor{\Upsilon}{^{\sigma)}}\tensor{\Upsilon}{^\nu}\right)
+2\tensor{g}{^\nu^{(\sigma}}\left(\tensor{R}{^{\rho)}^\mu}
-\tensor{\Upsilon}{^{\rho)}}\tensor{\Upsilon}{^\mu}\right)
-\tensor{g}{^\mu^\nu}\left(\tensor*{R}{^\rho^\sigma}+\tensor{\Upsilon}{^{\rho}^{;\sigma}}\right)\,,\label{AppCoefB2}\\
\tensor*{B}{^\mu^\nu^\rho_3}={}&\tensor{g}{^\rho^\beta}\left(\tensor{\Upsilon}{^\alpha}\tensor{P}{_{\alpha\beta}^{\mu\nu}}-2\tensor{P}{_{\alpha\beta}^{\mu\nu;\alpha}}\right)-\tensor{g}{^\alpha^\beta}\left(\tensor{\Upsilon}{^\rho}\tensor{P}{_{\alpha\beta}^{\mu\nu}}-2\tensor{P}{_{\alpha\beta}^{\mu\nu;\rho}}\right)\nonumber\\
&-\tensor{g}{^\rho^{(\nu}}\tensor{\Upsilon}{^{\mu)}}\left(\frac{f}{f_1}-R\right)
+\tensor{g}{^\mu^\nu}\left[\tensor{\Upsilon}{^\rho}\left(\frac{f}{f_1}-R\right)-2\tensor{\Upsilon}{^\alpha}\tensor*{R}{_{\alpha}^\rho}-\tensor{\Upsilon}{_\alpha^{;\alpha}^\rho}\right]\nonumber\\
&+\tensor{g}{^{\rho}^{(\nu}}\left(\tensor{\Upsilon}{_\alpha}\tensor*{R}{^{\mu)}^{\alpha}}+\tensor{R}{^{;\mu)}}+\tensor{\Upsilon}{_\alpha}^{;\alpha|\mu)}\right)-\tensor{\Upsilon}{^{(\mu}}\tensor*{R}{^{\nu)}^\rho}+4\,\tensor*{R}{^\rho^{(\mu}^{;\nu)}}\nonumber\\
&-2\tensor{\Upsilon}{_\alpha}\tensor{R}{^\alpha^{(\mu}^{\nu)}^\rho}-2 \tensor{W}{^\mu^\nu^{;\rho}}-3\tensor*{R}{^\mu^\nu^{;\rho}}\,,\label{AppCoefB3}\\
\tensor*{B}{^\mu^\nu_4}={}&-\left(\tensor{A}{^\alpha^\beta}\tensor{P}{_{\alpha\beta}^{\mu\nu}}\right)+\left(K\,\tensor{W}{^\mu^\nu}\right)+\tensor*{R}{^\alpha^{(\mu}}\left(\tensor*{R}{^{\nu)}_{\alpha}}-\tensor{\Upsilon}{^{\nu)}_{;\alpha}}\right)-\tensor{R}{^\mu_\alpha^\nu_\beta}\left(\tensor*{R}{^\alpha^\beta}-\tensor{\Upsilon}{^\alpha^{;\beta}}\right)\nonumber\\
&-\tensor{\Upsilon}{_\alpha}\left(\tensor*{R}{^\mu^\nu^{;\alpha}}-\tensor*{R}{^\alpha^{(\mu}^{;\nu)}}\right)+\Delta\tensor*{R}{^\mu^\nu}-\frac{1}{2}\tensor{\Upsilon}{^{(\mu}}\tensor{R}{^{;\nu)}}+\tensor{R}{^{;(\mu\nu)}}\,.\label{AppCoefB4}
\end{align}

\subsection{Coefficients of the operator $\mathbf{X}$}\label{AppCoeffX}

The coefficients of the operator $\mathbf{X}$, defined in \eqref{OpX}, are listed below explicitly. Since formally $\mathbf{X}^{\dagger}=\mathbf{X}$, the coefficient $\tensor*{X}{_3^{\mu\nu\rho}}$ can only be built from derivatives of the leading coefficient $\tensor*{X}{_2^{\mu\nu}}$. Without loss of generality, we have defined the $X_i$ as totally symmetric tensors. The explicit coefficients read  
\begin{align}
\tensor*{X}{_2^{\mu\nu}}={}&\tensor{g}{^\mu^\nu}\left(\frac{4}{3}R-\frac{f}{f_1}-\frac{1}{3}\frac{f_1}{f_2}+\frac{3}{4}\tensor{\Upsilon}{_\alpha}\tensor{\Upsilon}{^\alpha}-\frac{1}{2}\tensor{\Upsilon}{_\alpha^{;\alpha}}\right)+2\tensor*{R}{^\mu^\nu}-3\tensor{\Upsilon}{^\mu}\tensor{\Upsilon}{^\nu}+2\tensor{\Upsilon}{^{\mu;\nu}}\,,\label{AppCoefX2}\\
\tensor*{X}{_3^{\mu\nu\rho}}={}&-\tensor{X}{_2^{(\mu\nu;\rho)}}-\tensor{X}{_2^{\sigma(\mu}_{;\sigma}}\tensor{g}{^{\nu\rho)}}\,,\label{AppCoefX3}\\
\tensor*{X}{_4^{\mu\nu}}={}&\tensor{g}{^\mu^\nu}\left[-\frac{2}{3}\frac{f}{f_2}+\frac{2}{3}R\left(\frac{f_0}{f_1}+\frac{f_1}{f_2}+\tensor{\Upsilon}{_\alpha^{;\alpha}}-3\tensor{\Upsilon}{_\alpha}\tensor{\Upsilon}{^\alpha}-\frac{1}{2} R \right)+\tensor*{R}{_\alpha_\beta}\left(\frac{19}{6}\tensor{\Upsilon}{^\alpha}\tensor{\Upsilon}{^\beta}-\frac{2}{3}\tensor*{R}{^\alpha^\beta}\right)\nonumber\right.\\
&\left.-\frac{f}{f_1}\left(\frac{5}{6}\tensor{\Upsilon}{_\alpha}\tensor{\Upsilon}{^\alpha}+\tensor{\Upsilon}{_\alpha^{;\alpha}}\right)+\left(\tensor{\Upsilon}{_\alpha}\tensor{\Upsilon}{^\alpha}\right)\left(\frac{5}{16}\tensor{\Upsilon}{_\beta}\tensor{\Upsilon}{^\beta}+\frac{17}{6}\tensor{\Upsilon}{_\beta^{;\beta}}\right)-\frac{13}{12}\left(\tensor{\Upsilon}{_\alpha^{;\alpha}}\right)^2\right]\nonumber\\
&+\tensor*{R}{^\mu^\nu}\left(\frac{2}{3}\frac{f}{f_1}+\frac{13}{6}\tensor{\Upsilon}{_\alpha}\tensor{\Upsilon}{^\alpha}-\frac{8}{3}R+4\tensor{\Upsilon}{_\alpha^{;\alpha}}\right)+\tensor{R}{^\mu^\alpha^\nu^\beta}\left(\frac{4}{3}\tensor*{R}{_\alpha_\beta}-\frac{14}{4}\tensor{\Upsilon}{_\alpha}\tensor{\Upsilon}{_\beta}\right)\nonumber\\
&+\tensor{\Upsilon}{^\mu}\tensor{\Upsilon}{^\nu}\left(\frac{19}{3}\frac{f}{f_1}-\frac{5}{4}\tensor{\Upsilon}{_\alpha}\tensor{\Upsilon}{^\alpha}+\frac{10}{3}R-\frac{31}{6}\tensor{\Upsilon}{_\alpha^{;\alpha}}\right)+\tensor{\Upsilon}{^{\mu;\nu}}\left(-2\frac{f}{f_1}-\frac{1}{3}R+\frac{5}{6}\tensor{\Upsilon}{_\alpha}\tensor{\Upsilon}{^\alpha}+\frac{7}{3}\tensor{\Upsilon}{_\alpha^{;\alpha}}\right)\nonumber\\
&+\frac{4}{3}\tensor*{R}{^\mu^\alpha}\tensor*{R}{_\alpha^\nu}-\frac{10}{3}\tensor*{R}{_\alpha^{(\mu}}\tensor{\Upsilon}{^{\nu)}}\tensor{\Upsilon}{_\alpha}+\text{t.d.}\,,\label{AppCoefX4}
\end{align}
where we have neglected total derivative terms in $\tensor*{X}{_4^{\mu\nu}}$.

\section{Representations of the final result}\label{App:RepFinRes}

\subsection{Representation in terms of $E$}
For the on shell reduction, it is convenient to express the final result in terms of the rescaled extremal $\tensor{E}{_\mu_\nu}$ and its trace $E$.
This representation makes the terms which are proportional to the equations of motion manifest.
The conversion from the $\Upsilon$-representation \eqref{Gamma1LoopOffShellFinalGamPara} is performed by a systematic procedure which involves several integration by parts identities listed below.
The procedure involves the following steps.
First, we eliminate structures that contain derivatives of $\tensor{\Upsilon}{_\mu}$ in three steps.
Structures with three derivatives $\nabla\nabla\nabla\Upsilon$ are total divergences and are neglected.
Structures which involve second derivatives $\Upsilon\nabla\nabla\Upsilon$ are partially integrated to structures that only involve first-order derivatives $\nabla\Upsilon\nabla\Upsilon$.
Finally, structures that involve first-order derivatives $\nabla\Upsilon$ are converted into structures that involve the rescaled extremal, curvatures and $\Upsilon\Upsilon$ structures.
In this way, all structures involving derivatives of $\Upsilon$ are eliminated.
Finally, structures that involve undifferentiated $\Upsilon$ terms are eliminated by the integration by parts identities we list below.
The equalities $\overset{\bullet}{=}$ are to be understood to hold only under the integral sign modulo surface terms:
\begin{align}
\tensor{\Upsilon}{^{\mu;\nu}}&=\tensor{E}{^\mu^\nu}-\frac{1}{3}\left(E+R-\frac{1}{2}\frac{f}{f_1}\right)\tensor{g}{^\mu^\nu}+\tensor*{R}{^\mu^\nu}-\tensor{\Upsilon}{^\mu}\tensor{\Upsilon}{^\nu}\,,\label{CDGamextremal}\\
(\tensor{\Upsilon}{_\mu}\tensor{\Upsilon}{^\mu})^2&\overset{\bullet}{=}-\frac{1}{3}\left(E+R-\frac{f}{f_1}\right)(\tensor{\Upsilon}{_\mu}\tensor{\Upsilon}{^\mu})+\frac{2}{3}\left(\tensor{E}{_\mu_\nu}+\tensor*{R}{_\mu_\nu}\right)\tensor{\Upsilon}{^\mu}\tensor{\Upsilon}{^\nu}\,,\\
\frac{f_1}{f_2}(\tensor{\Upsilon}{_\mu}\tensor{\Upsilon}{^\mu})&\overset{\bullet}{=}R\,\Delta\ln f_1\,,\\
\frac{f}{f_1}(\tensor{\Upsilon}{_\mu}\tensor{\Upsilon}{^\mu})
&\overset{\bullet}{=}-\frac{1}{6}\left(E+R-2\frac{f}{f_1}\right)\frac{f}{f_1}+\frac{1}{2}R\,\Delta\ln f_1\,,\\
R\,(\tensor{\Upsilon}{_\mu}\tensor{\Upsilon}{^\mu})
&\overset{\bullet}{=}-\frac{1}{3}\left(E+R-2\frac{f}{f_1}\right)R+R\,\Delta\ln f_1 \,,\\
E\,(\tensor{\Upsilon}{_\mu}\tensor{\Upsilon}{^\mu})&\overset{\bullet}{=}-\frac{1}{3}\left(E+R-2\frac{f}{f_1}\right)E
+E\,\Delta\ln f_1 \,,\\
\tensor{E}{_\mu_\nu}\tensor{\Upsilon}{^\mu}\tensor{\Upsilon}{^\nu}
&\overset{\bullet}{=}\frac{1}{2}\tensor{E}{_\mu_\nu}\tensor{E}{^\mu^\nu}
+\frac{1}{2}\tensor{E}{_\mu_\nu}\tensor*{R}{^\mu^\nu}
-\frac{1}{6}E\left(E+R-\frac{1}{2}\frac{f}{f_1}\right)\,,\\
\tensor*{R}{_\mu_\nu}\tensor{\Upsilon}{^\mu}\tensor{\Upsilon}{^\nu}&\overset{\bullet}{=}\tensor{E}{_\mu_\nu}\tensor*{R}{^\mu^\nu}+\tensor*{R}{_\mu_\nu}\tensor*{R}{^\mu^\nu}-\frac{R}{3}\left(E+R-\frac{1}{2}\frac{f}{f_1}\right)+\frac{1}{2}R\,\Delta\ln f_1\,.
\end{align} 
Applying these rules we express the final result \eqref{Gamma1LoopOffShellFinalGamPara} in terms of the extremal $\tensor{E}{_\mu_\nu}$ and its trace $E$,
\begin{align}
\left. \varGamma_1 \right|^{\mathrm{div}}=\frac{1}{32\pi^2\varepsilon}\int{\rm d}^4x\, g^{\nicefrac{1}{2}}&\left[
-\frac{71}{60}\mathcal{G}
-\frac{609}{80}\tensor*{R}{_\mu_\nu}\tensor*{R}{^\mu^\nu}
+\frac{1}{3}\frac{f}{f_2}
-\frac{115}{288}\left(\frac{f}{f_1}\right)^2
-\frac{1}{18}\left(\frac{f_1}{f_2}\right)^2
-\frac{15}{64}\frac{f}{f_1}\,R\right.\nonumber\\
&\left.\;\,
+\frac{3919}{1440}R^2
+\frac{15}{64}R\,\Delta\ln f_1
+E\left(\frac{55}{108}E
-\frac{419}{432}\frac{f}{f_1}
+\frac{2933}{864}R
+\frac{221}{288}\Delta\ln f_1\right)\right.\nonumber\\
&\left.\;\,-\tensor{E}{_\mu_\nu}\left(\frac{403}{96}\tensor{E}{^\mu^\nu}
+\frac{2987}{288}\tensor*{R}{^\mu^\nu}\right)\right]\,.\label{Gamma1LoopOffShellFinalEPara}
\end{align}

\subsection{Representation in terms of $\nabla R$}
The final result can be expressed in terms of the Ricci scalar $R$ and its derivatives. Moreover, the quadratic curvature invariants can be represented in terms of the Ricci scalar $R$, the Weyl tensor $\tensor{C}{_\mu_\nu_\rho_\sigma}$ and the trace-free Ricci tensor $\tensor*{S}{_\mu_\nu}$. In four dimensions the latter are defined as
\begin{align}
\tensor{C}{_\mu_\nu_\rho_\sigma}\coloneqq{}&\tensor{R}{_\mu_\nu_\rho_\sigma}-\left(\tensor*{R}{_\mu_{[\rho}}\tensor{g}{_{\sigma]}_\nu}+\tensor*{R}{_\nu_{[\sigma}}\tensor{g}{_{\rho]}_\mu}\right)+\frac{1}{3}R\,\tensor{g}{_\mu_{[\rho}}\tensor{g}{_{\sigma]}_\nu}\label{Weyl}\,,\\
\tensor*{S}{_\mu_\nu}\coloneqq{}&\tensor*{R}{_\mu_\nu}-\frac{1}{4}\tensor{g}{_\mu_\nu}R\label{TFRicci}\,.
\end{align}
This representation is best suited for the reduction to constant curvature backgrounds as the Weyl tensor $\tensor{C}{_{\mu\nu\rho\sigma}}$ and the trace-free Ricci tensor $\tensor*{S}{_{\mu\nu}}$ vanish in this case:
\begin{align}
\tensor{C}{_{\mu\nu\rho\sigma}}={}&0\,,\quad
\tensor*{S}{_{\mu\nu}}={}0\,.
\end{align}
The result \eqref{Gamma1LoopOffShellFinalGamPara}, expressed in terms of the Ricci scalar and the trace-free contributions to Riemann tensor, takes the form
\begin{align}
\left. \varGamma_1 \right|^{\mathrm{div}}={}\frac{1}{32\pi^2\varepsilon}\int{\rm d}^4x\, g^{\nicefrac{1}{2}}&\left\{
-\frac{9}{2}\left(\frac{f}{f_1}\right)^2
-\frac{1}{18}\left(\frac{f_1}{f_2}\right)^2
+\frac{9}{2}\frac{f}{f_1}R
+\frac{1}{3}\frac{f}{f_2}
-\frac{173}{240}R^2
+\frac{167}{180}\tensor{S}{_\mu_\nu}\tensor{S}{^\mu^\nu}
-\frac{71}{60}\tensor{C}{_\mu_\nu_\alpha_\beta}\tensor{C}{^\mu^\nu^\alpha^\beta}
\right.\nonumber\\
&\;\,\left.
+\frac{20}{9}\left(\frac{f_2}{f_1}\right)^2\tensor{S}{_\mu_\nu}\tensor{R}{^{;\mu}}\tensor{R}{^{;\nu}}
+\left[\frac{69}{4}\frac{f}{f_2}\left(\frac{f_2}{f_1}\right)^3
-\frac{9}{4}\frac{f_2}{f_1}
-\frac{331}{57}\left(\frac{f_2}{f_1}\right)^2R\right]\tensor{R}{^{;\mu}}\tensor{R}{_{;\mu}}
\right.\nonumber\\
&\;\,\left.
+\left[\frac{137}{12}\frac{f_3}{f_2}\left(\frac{f_2}{f_1}\right)^2
-\frac{247}{24}\left(\frac{f_2}{f_1}\right)^3\right]\left(\tensor{R}{^{;\mu}}\tensor{R}{_{;\mu}}\right)\Delta R
-\frac{137}{24}\left(\frac{f_2}{f_1}\right)^2(\Delta R)^2\right.\nonumber\\
&\;\,\left.
+\left[\frac{247}{24}\frac{f_3}{f_2}\left(\frac{f_2}{f_1}\right)^3
-\frac{137}{24}\left(\frac{f_3}{f_1}\right)^2
-\frac{403}{32}\left(\frac{f_2}{f_1}\right)^4
\right]\left(\tensor{R}{^{;\mu}}\tensor{R}{_{;\mu}}
\right)^2
\right\}\,.\label{Gamma1LoopOffShellFinalRPara}
\end{align}

\section{Functional traces for constrained fields}\label{TracesIrredDec}
The traces in Sec. \eqref{SubSec:IrredDec} are taken over fields which satisfy the constraints \eqref{DiffConstraint}; that is, we have to evaluate traces over transversal-traceless tensor fields and transversal vector fields. 
The constrained heat traces are given by (see e.g.\ the third and sixth column in Table 9 of \cite{Codello:2008vh})
\begin{align}
\operatorname{Tr}_{1,\perp}e^{-s\Delta}&=\frac{1}{(4\pi s)^{2}}\left[3+\frac{1}{4}R_0s-\frac{7}{1440}R_0^2s^2+{\cal O}(s^3)\right]\,,\\
\operatorname{Tr}_{2,\perp}e^{-s\Delta}&=\frac{1}{(4\pi s)^{2}}\left[5-\frac{5}{6}R_0\,s-\frac{1}{432}R_0^2s^2+{\cal O}(s^3)\right]\,.
\end{align}
With these expressions, we calculate the divergent contributions of the trace over constrained fields for a generalized Laplacian $\Delta+P$ with constant scalar potential $P$,
\begin{align}
\left. \operatorname{Tr}_{2,\perp}\ln\left(\Delta+P\right) \right|^{\mathrm{div}}&=\frac{1}{\varepsilon}\left[\frac{1}{18}-20\frac{P}{R_0}-60\left(\frac{P}{R_0}\right)^2\right]\,,\\
\left. \operatorname{Tr}_{1,\perp}\ln\left(\Delta+P\right) \right|^{\mathrm{div}}&=\frac{1}{\varepsilon}\left[\frac{7}{60}+6\frac{P}{R_0}-36\left(\frac{P}{R_0}\right)^2\right]\,.
\end{align}
This allows us to extract the divergences in \eqref{TTTensorTrace} and \eqref{PrimeTrVector}. 
Almost all of the remaining scalar traces are of the standard form for minimal second-order operators, such that we directly employ \eqref{TRLNDiv} to evaluate them.
The only nonstandard trace is that of the fourth-order scalar operator. Expanding the expression around the principal part $\Delta^2$ and using the universal functional traces \eqref{UFTs}, we find  
\begin{align}
\left. \operatorname{Tr}_0\ln\left[\Delta^2+\frac{1}{3}\left(\frac{f_1}{f_2}-\frac{5}{2}R_0\right)\Delta+\frac{1}{3}\left(\frac{f}{f_2}-\frac{f_1}{f_2}R_0+\frac{1}{2}R_0^2\right)\right] \right|^{\mathrm{div}}&{}=\frac{1}{\varepsilon}\left[-\frac{374}{45}-\frac{4}{3}\left(\frac{f_1}{f_2R_0}\right)^2+8\frac{f}{f_2R_0^2}\right]\,.
\end{align}

\twocolumngrid

\bibliography{HKfRV2}{}

\end{document}